\documentclass[11pt, jpg]{iopart}
\usepackage{iopams}
\expandafter\let\csname equation*\endcsname\relax
\expandafter\let\csname endequation*\endcsname\relax
\usepackage{mathtools}
\usepackage{import}
\usepackage{graphicx}
\usepackage{fancyhdr}
\usepackage{tikz}
\usepackage{array}
\usepackage{courier}
\usepackage{slashed}
\usepackage{bm}
\usepackage{dsfont}
\usepackage{epigraph}
\usepackage{verbatim}
\usepackage{listings}
\usepackage{comment}
\usepackage{multirow}
\usepackage{wrapfig} 
\usepackage{hyperref}
\usepackage{enumerate}

\usepackage[textsize=normalsize]{todonotes}

\DeclarePairedDelimiter\ceil{\lceil}{\rceil}

\newcommand*\diff{\mathop{}\!\mathrm{d}}

\newcommand{\NWP}{$N_{\text{WP}}$}
\newcommand{\SWAVE}{$^1\text{S}_0$}
\newcommand{\STWAVE}{$^3\text{S}_1$}
\newcommand{\NN}{NN}

\newcommand{\np}{np}
\newcommand{\pp}{pp}
\newcommand{\wpcd}{WPCD}
\newcommand{\ls}{LS}

\newcommand{\cheft}{$\chi$EFT}
\newcommand{\mi}{MI}
\newcommand{\swp}{SWP}
\newcommand{\fwp}{FWP}

\newcommand\inputpgf[2]{{
		\let\pgfimageWithoutPath\pgfimage
		\renewcommand{\pgfimage}[2][]{\pgfimageWithoutPath[##1]{#1/##2}}
		\input{#1/#2}
}}

\usetikzlibrary{shapes,arrows}	
\usetikzlibrary{arrows,decorations.markings}
\usetikzlibrary{positioning}
\definecolor{decisionColor}{HTML}{DEE1B2}
\definecolor{nodeColor}{HTML}{EBFAFF}
\tikzstyle{roundrect}=[rectangle, rounded corners, minimum width=0cm, minimum height=0.5cm, text centered, draw=black, fill=nodeColor, text width=2.2cm] 
\tikzstyle{decision} = [diamond, minimum width=3.7cm, minimum height=3.2cm, text centered, draw=black, fill=decisionColor, scale=1, text width=2.05cm]
\tikzstyle{arrow} = [decoration={markings,mark=at position 1 with
	{\arrow[scale=1.5,>=stealth]{>}}},postaction={decorate}]
\tikzset{
	desicion/.style={
		diamond,
		draw,
		text width=3em,
		text badly centered,
		inner sep=0pt
	},
	block/.style={
		rectangle,
		draw,
		text width=10em,
		text centered,
		rounded corners
	},
	cloud/.style={
		draw,
		ellipse,
		minimum height=2em
	},
	descr/.style={
		fill=white,
		inner sep=2.5pt
	},
	connector/.style={
		-latex
	},
	rectangle connector/.style={
		connector,
		to path={(\tikztostart) -- ++(#1,0pt) \tikztonodes |- (\tikztotarget) },
		pos=0.5
	},
	rectangle connector/.default=-2cm,
	straight connector/.style={
		connector,
		to path=--(\tikztotarget) \tikztonodes
	}
}

\begin{document}
\title{Wave-packet continuum discretisation for nucleon-nucleon scattering predictions}

\date{\today}
\author{Sean B. S. Miller, Andreas Ekstr\"om, Christian Forss\'en} 
\address{Department of Physics, Chalmers University of Technology, SE-412 96 Gothenburg, Sweden}
\eads{\mailto{sean.miller@chalmers.se}, \mailto{andreas.ekstrom@chalmers.se}}

\begin{abstract} 
  In this paper we analyse the efficiency, precision, and accuracy of
  computing elastic nucleon-nucleon (\NN) scattering amplitudes with
  the wave-packet continuum discretisation method (\wpcd). This method
  provides approximate scattering solutions at multiple scattering
  energies simultaneously. We therefore utilise a graphics processing
  unit (GPU) to explore the benefits of this inherent
  parallelism. From a theoretical perspective, the \wpcd{} method
  promises a speedup compared to a standard matrix-inversion
  method. We use the chiral NNLO$_{\rm opt}$ interaction to
  demonstrate that \wpcd{} enables efficient computation of \NN{}
  scattering amplitudes provided one can tolerate an averaged method
  error of $~1-5$ mb in the total cross section at scattering energies
  $0-350$ MeV in the laboratory frame of reference. Considering only
  scattering energies $\sim40-350$ MeV, we
  find a smaller method error of $\lesssim 1-2$ mb. By increasing
  the number of wave-packets we can further reduce the overall method
  error. However, the parallel leverage of the \wpcd{} method will be
  offset by the increased size of the resulting discretisation
  mesh. In practice, a GPU-implementation is mainly advantageous for
  matrices that fit in the fast on-chip shared memory. We find that
  \wpcd{} is a promising method for computationally efficient,
  statistical analyses of nuclear interactions from effective
  field theory, where we can utilise Bayesian inference methods
  to incorporate relevant uncertainties.
\end{abstract}


\submitto{\jpg}

\maketitle
\section{Introduction}
A large portion of interaction-potential models currently applied in
\textit{ab initio} many-nucleon calculations are constructed using
ideas from chiral effective field theory
(\cheft)~\cite{doi:10.1146/annurev.nucl.52.050102.090637,RevModPhys.81.1773,MACHLEIDT20111,RevModPhys.92.025004}.
Such potentials typically contain $\gtrsim$10-30 low-energy constants
(LECs), acting as physical calibration parameters, that must be
inferred from data. Bayesian methods for parameter estimation offer
several advantages in this regard, in particular in conjunction with
EFTs, see e.g. Ref.~\cite{Wesolowski_2016}. However, making reliable
inferences typically incur a high total computational cost due to the
large amount of posterior samples. It is therefore important to
establish an efficient computational framework for generating model
predictions of physical observables.

In this paper we look to low-energy nucleon-nucleon (\NN) scattering
cross sections as it constitutes the bulk part of the standard dataset
for inferring the most probable values of the LECs, see
e.g. Refs.~\cite{PhysRevC.68.041001,PhysRevX.6.011019,Reinert:2018tb,PhysRevC.91.024003}.
The most recent, and statistically consistent,
database~\cite{PhysRevC.88.064002} of \NN{} scattering cross sections
contains data for thousands of measured proton-proton (\pp) and
neutron-proton (\np) cross sections at hundreds of different
laboratory scattering energies, mostly below the pion-production
threshold at $T_{\text{lab}} \approx 290$ MeV. In most cases, the computational bottleneck
when predicting \NN{} scattering amplitudes for a given interaction potential comes from
obtaining numerical solutions to the Lippmann-Schwinger (\ls{}) equation.

There are essentially three approaches for improving computational
efficiency and speed of a computational procedure: i) develop improved
numerical methods and algorithms tailored to the physical model at
hand and its application, ii) use specific hardware, e.g. a faster
CPU, increased memory bandwidth, or parallel architectures such as in
a graphics processing unit (GPU) to better handle some of the most
dominant computational procedures of the model, or iii) replace any
computationally expensive model evaluations with a fast, as well as
sufficiently accurate and precise, surrogate model, i.e. an
\textit{emulator}, which mimics the original model output. This latter
approach is very interesting and in particular eigenvector
continuation (EC)~\cite{Frame:2017fah,Konig:2019adq} applied to
emulate \NN{}-scattering
amplitudes~\cite{Furnstahl:2020abp,Melendez:2021lyq} shows great
promise, although uncertainty quantification is yet to be explored.
Note, however, that EC emulation attains most of its speedup when
applied to potential models that exhibit linear parameter
dependencies. In cases where such dependencies are not present one
might resort to other methods to handle the non-linear response, such
as EC combined with Gaussian process emulation
~\cite{zhang2021fast}. However, one should note that Gaussian process
emulation~\cite{10.5555/1162254} and other standard machine learning
methods exhibit poor scaling with increasing dimensionality of the
input parameter domain.

Here, we will focus on approaches i) and ii) by exploring both the
standard matrix-inversion (\mi{}) method, see e.g.
Ref.~\cite{Haftel1970}, and the wave-packet continuum discretisation
(\wpcd{}) method~\cite{RUBTSOVA2015613} for solving the \ls{}
equation.  The \wpcd{} method basically corresponds to a bound-state
approach that uses eigenfunctions of the full \NN{} Hamiltonian to
approximate scattering solutions at any on-shell energy.  This method
is particularly interesting since it provides approximate scattering
amplitudes at multiple scattering energies simultaneously. We have
therefore implemented this inherently parallel method on a GPU. We
also note that \wpcd{} places no constraint on the analytical form of
the potential or its parametric dependence. As such, \wpcd{}
acceleration for \NN{} scattering complements the EC approaches for
emulation~\cite{Furnstahl:2020abp,Melendez:2021lyq,zhang2021fast}
mentioned above.

Speeding up computations very often comes at the expense of accuracy
and/or precision, and the \wpcd{} method is no exception to this
principle. The magnitude of experimental errors in the calibration
data and the estimated theoretical
model-discrepancies~\cite{Brynjarsdottir:2014cb} provide natural
tolerances for the level of method error that is acceptable. Indeed it
is undesirable to have a method error that dominates the error budget
such that it obscures or even hampers the inference of useful
information. We therefore analyse the \wpcd{} method in detail and
quantify realistic computational speedups and compare the method
errors to recent estimates of the model discrepancy in
\cheft~\cite{PhysRevC.96.024003}. We also analyse and compare the
numerical complexities of the \mi{} and \wpcd{} methods.

\section{Nucleon-Nucleon Scattering}
The \ls{} equation, in operator form, for the transition-matrix
operator $\hat{T}$ at some scattering energy $E$ is given by
\begin{equation}
	\hat{T}(E) = \hat{v} + \hat{v}\hat{g}_0(E)\hat{T}.
	\label{eq:LS}
\end{equation}
This is an inhomogeneous Fredholm integral equation where $\hat{v}$ is
some \NN{}-potential operator, $\hat{g}_0(E) = (E-\hat{h}_0
+i\epsilon)^{-1}$ is the free Green's function, and $\hat{h}_{0}$ is
the free Hamiltonian, i.e. the kinetic energy operator, and $i\epsilon
\rightarrow0$ is the positive imaginary part of the complex
energy. Most realistic \NN{} potentials in nuclear \textit{ab initio}
calculations do not furnish analytical solutions for the
$T$-matrix. It is however straightforward to numerically solve for the
$T$-matrix for an on-shell energy $E$. There exists e.g. variational
methods~\cite{lh1,lh2,PhysRev.74.1763,PhysRev.141.1468}, as well as
Neumann or Born series expansions for sufficiently weak
potentials. One can also try Pad\'e
extrapolants~\cite{doi:10.1002/prop.19720200502} in cases where the
integral kernel is not sufficiently perturbative to converge the
resulting Neumann series. In this work we employ the standard \mi{}
method~\cite{Haftel1970} which amounts to inverting a relatively small
complex-valued matrix at each scattering energy. This is a trivial
operation that can be straightforwardly carried out within
milliseconds on a modern CPU. However, solving at multiple scattering
energies to obtain all cross-sections present in the \NN{} database
amounts to at least a few seconds of computation. In a Bayesian
analysis, where one repeatedly evaluates a likelihood function across
a multi-dimensional parameter domain, any speedup in the solution of
the \ls{} equation will directly impact the total computation time.

The momentum-space partial-wave representation of the \ls{} equation
for the \NN{} $T$-matrix, Eq.~\eqref{eq:LS}, is given by
\begin{align}
  \begin{split}
    T^{sJ}_{l'l}(q',q;E) = {v}^{sJ}_{l'l}(q',q) + 
    \frac{2}{\pi}\sum_{\l''}&\int_0^\infty dk \: k^2{v}^{sJ}_{l'l''}(q',k)\\ \times &g_0(k;E)
    T^{sJ}_{l''l}(k,q;E)\:,
    \label{eq | LS equation in momentum representation without angles}
  \end{split}
\end{align}
where $q$, $q'$, and $k$ are relative momenta, $E=p^2/m_N$ is the
centre-of-mass (c.m.) energy with momentum $p$, $m_N$ is the nucleon
mass, and $g_0(k;E)\equiv(E-k^2/m_N+i\epsilon)^{-1}$. In this work we
only consider the canonical \NN{} interaction potential, and therefore
use the shorthand notation $T^{sJ}_{l'l}(q',q) \equiv \langle
q',l',s,J|\hat{T}|q,l,s,J\rangle$ for partial-wave
amplitudes. Furthermore we use the following normalisation of momentum
states, $\langle k'|k\rangle = \frac{\delta(k'-k)}{k'k}$.  We suppress
isospin notation since the total isospin $T$ is defined uniquely by
the Pauli principle given the spin and angular momentum
quantum numbers $s$ and $l$, respectively, while the projected isospin
$T_z$ is defined at the
outset of, and conserved throughout, a scattering observable
calculation.  Methods for obtaining the partial-wave-projected
potential $v^{sJ}_{l'l}(q',q)$ can be found in
e.g. Ref.~\cite{Erkelenz:1971caz}.  Integrated and differential
scattering cross sections at a specific scattering energy $E$ can be
straightforwardly evaluated given the partial-wave $T$-matrix using
the expressions presented in~\ref{app | scattering theory}.

Throughout this work we will compare the efficiency and accuracy of
the \wpcd{} method, presented in Sec.~\ref{sec:WPCD}, to a set of
numerically exact results obtained using the \mi{} method presented
below.

\subsection{Matrix-inversion method for solving the Lippmann-Schwinger equation\label{sec:MI}}
For the resolvent $\hat{g}_{0}(E)$, the kernel in the
\ls{} equation \eqref{eq | LS equation in momentum representation
  without angles} has a pole singularity at $k = p$. This can be
handled via a principal-value decomposition\footnote{Kramers-Konig relation, dispersion relation, or the
  Sokhotski-Plemelj identity.}, i.e.
\begin{equation}
  \underset{\epsilon \rightarrow 0}{\rm lim}\:\frac{1}{x \pm i\epsilon} = \mathcal{P}\left(\frac{1}{x}\right)  \mp i \pi \delta(x) \:,\quad x\in\mathds{R},
  \label{eq:kk}
\end{equation}
such that the remaining integral can be evaluated using
e.g. Gauss-Legendre quadrature on some grid $\{k_i\}_{i=1}^{N_{Q}}$ of
momenta $k_i\neq p$ with corresponding weights
$w_i$. Following~\cite{Haftel1970}, the complex $T$-matrix in
Eq.~\eqref{eq | LS equation in momentum representation without angles}
can be solved via the inversion of a finite-dimensional matrix
equation for the on-shell momentum $q=p=\sqrt{m_NE}$,
\begin{align}
  \begin{split}
    T_{l'l}^{sJ}(q,q;E) = v_{l'l}^{sJ}(q,q) +
    \frac{2}{\pi}\sum_{l''}&\sum_{i=1}^{N_Q} w_iv_{l'l''}^{sJ}(q,k_i)\\ &\times g_0(k_i;E)T_{l''l}^{sJ}(k_i,q)\:.
  \end{split}
  \label{eq | LS equation in operator form}
\end{align}
We can introduce a basis $q_i\in\{k_1,k_2,\ldots, k_{N_Q}, p\}$ such
that the operators can be written in matrix form with row $i$ and column $j$
corresponding to $q_i$ and $q_j$ respectively, i.e. $(V_{l'l}^{sJ})_{ij} \equiv \langle
q_i|v_{l'l}^{sJ}|q_j\rangle$. The on-shell $T$-matrix element
is then given by $(T_{l'l}^{sJ})_{N_Q+1,N_Q+1}$. Introducing a vector $D$
with elements defined as
\begin{equation}
  D_{i} \equiv \begin{cases} \frac{2w_ik_i^2m_N}{\pi (k_i^2 - p^2)}
    \quad &\text{if}\quad i\leq N_Q
    \:,\\ -\sum_{i=1}^{N_Q}\frac{2w_ip^2m_N}{\pi (k_i^2 - p^2)} +
    \frac{i\pi pm_N}{2}\quad &\text{if}\quad i=N_Q+1 \:,
  \end{cases}
\end{equation}
Eq.~\eqref{eq | LS equation in operator form} can be rewritten as a linear system of equations,
\begin{equation}
  \sum_{l''}F_{l'l''}^{sJ} T_{l''l}^{sJ} = V_{l'l}^{sJ} \:,
  \label{eq | LS equation in matrix form with wave matrix}
\end{equation}
where we have introduced the wave matrix $F_{l'l}^{sJ}$,
\begin{equation}
  (F_{l'l}^{sJ})_{ij} \equiv \delta_{ij}\delta_{l'l} - (V_{l'l}^{sJ})_{ij}D_j \:.
  \label{eq | definition wave matrix}
\end{equation}

Direct inversion of $F_{l'l}^{sJ}$ in Eq.~\eqref{eq | LS equation in
  matrix form with wave matrix} is usually discouraged in scientific
computing due to the instability of matrix inversion algorithms
\cite{10.1093/imanum/12.1.1}. A more advisable practice is to use
LU-decomposition. Additionally, it is numerically more stable to first
introduce the $K$-matrix\footnote{Also referred to as the reactance
  $R$-matrix.} as the principal value part of the \ls{}
equation. Since the potential $v$ is real, the $K$-matrix is also
real. This leads to a set of purely real
matrix-equations~\cite{Haftel1970} based on a nonsingular integral.

The on-shell momentum dependence in $F_{l'l}^{sJ}$ demands the solution of
an entire linear system of equations for every energy of
interest. Formally, the $T$-matrix is defined by the potential
operator via \cite{Glockle99109}
\begin{equation}
  T^{sJ}_{l'l}(q',q) \equiv \langle q' | \hat{v}^{sJ}_{l'l} |\psi_q^+\rangle \:,
  \label{eq | T-matrix definition}
\end{equation}
where $|\psi^+_q\rangle$ are eigenstates (outbound scattering states)
of the full Hamiltonian $\hat{h}$ with momentum $q$. With this, we can instead
express the \ls{} equation using the full resolvent $\hat{g}(E) =
(E-\hat{h}+ i\epsilon)^{-1}$,
\begin{align}
  \begin{split}
    T^{sJ}_{l'l}(q',q;E) = v^{sJ}_{l'l}(q',q) + \frac{2}{\pi}\sum_{\l''}\int_0^\infty
    &dk \:k^2 v^{sJ}_{l'l''}(q',\psi^+_k)\\ \times &g(k;E)v^{sJ}_{l''l}(\psi^+_k,q)\:,
  \end{split}
  \label{eq | LS equation with full resolvent}
\end{align}
where $v^{sJ}_{l'l}(q',\psi^+_k)=\langle
q'|\hat{v}^{sJ}_{l'l}|\psi^+_k\rangle$ and $g(k;E) = (E-k^2/m_N+
i\epsilon)^{-1}$. This is significantly easier to evaluate numerically
using quadrature since it only amounts to matrix
multiplications. However, the scattering states $|\psi_q^+\rangle$ are
not available at the outset. This is where the \wpcd{} method enters
to effectively approximate the scattering states using
square-integrable eigenstates of the \NN{} Hamiltonian.

\section{Wave-packet continuum discretisation}
The \wpcd{} method~\cite{RUBTSOVA2015613} effectively eliminates the
requirement to explicitly solve the \ls{} equation at several
scattering energies $E$ using matrix inversion. Instead, one
diagonalises the full Hamiltonian in a finite basis, and uses the
resulting discrete set of eigenstates to approximate all scattering
states of interest. Equipped with these states, this approach enables
straightforward evaluation of the \emph{full} resolvent at any value
of the on-shell scattering energy $E$, which makes the \wpcd{} method
intrinsically parallel with respect to obtaining scattering solutions
at different energies. This is one of several known bound-state
techniques to solve the multi-particle scattering
problem~\cite{CARBONELL201455}. To provide a self-contained
presentation, we devote this section to introduce the \wpcd{} method
for describing elastic \NN{} scattering, starting with a definition of
a finite wave-packet basis.
	
\subsection{Scattering observables in a finite basis}
Generally we can project some Hamiltonian state $|\Psi(E)\rangle$ with
positive energy $E$ onto a complete basis (including both bound and
free basis states). In this case, the expectation value of an operator
$\hat{O}(\hat{h})$ depending purely on the full Hamiltonian $\hat{h}$
can be represented in the following form,
\begin{align}
	\begin{split}
	\langle \Psi(E)|\hat{O}|\Psi(E)\rangle =
	&\sum_{i=1}^{n_b}u(\epsilon_i^b)|\langle\Psi|\psi_i^b\rangle|^2
	\\ &+\int_0^\infty \diff E'
	u(E')|\langle\Psi|\psi(E')\rangle|^2 \:,
	\end{split}
	\label{eq | operator in continuous basis}
\end{align}
where $\{|\psi_i^b\rangle\}_{i=1}^{n_b}$ are bound states with
energies $\epsilon_i^b$ and $|\psi(E')\rangle$ are free states with
energy $E'$, both of which are eigenstates of $\hat{h}$, and we have
defined $u(\epsilon_i^b)=\langle\psi_i^b|\hat{O}|\psi_i^b\rangle$ and
$u(E')=\langle \psi(E')|\hat{O}|\psi(E')\rangle$. Naturally, we have
$|\Psi(E)\rangle=|\psi(E)\rangle$ (for $E>0$) as $\hat{h}$ is the
Hamiltonian and $\Psi$ the eigenstate, collapsing the expansion
above.  However, this identity is not useful in numerical approaches
since we do not know $\psi(E)$, leaving us to solve the integral in a
finite, approximative basis.

A computational routine
for evaluating integrals, such as quadrature, uses a finite
mesh of points where the integrand is evaluated. In scattering, this
approach requires a finite basis of states $|\widetilde{\psi}_i\rangle$
with corresponding positive energies $E_i$. These states do not span 
the whole continuous momentum-space and are thus referred to as pseudostates.
Below, we will demonstrate how to construct the pseudostates in
a wave-packet basis. The pseudostates form a basis for a
finite quadrature-prescription to express an operator
\begin{equation}
  \langle \Psi|\hat{O}(\hat{h})|\Psi\rangle \approx \sum_{i=1}^{n_b}u(\epsilon_i^b)|\langle\Psi|\psi_i^b\rangle|^2 + \sum_{i=1}^{n} u(E_i)|\langle\Psi|\widetilde{\psi}_i\rangle|^2 \:,
  \label{eq | oeprator in discrete basis}
\end{equation}
where we introduce quadrature weights $w_i$ such that
\begin{equation}
  |\langle\Psi|\widetilde{\psi}_i\rangle|^2 = w_i|\langle\Psi|\psi(E_i)\rangle|^2 \:.
  \label{eq | pseudostate to cont. state identity}
\end{equation}
To determine the weights $w_i$ we can use an equivalent quadrature
(EQ) technique
\cite{Heller19751222,Winick1978910,Winick1978925,Heller19732946,Corcoran1977}
in which it can be shown that the weights do not depend on the state
$|\Psi\rangle$. The weights represent a type of transformation
coefficient between pseudostates $|\widetilde{\psi}_i\rangle$ and 
the states $|\psi(E')\rangle$.
An approximate relation based on
Eq.~\eqref{eq | pseudostate to cont. state identity} can then be
introduced,
\begin{equation}
  \langle \psi(E_i)|\hat{O}|\psi(E_i)\rangle \approx \frac{\langle \widetilde{\psi}_i|\hat{O}|\widetilde{\psi}_i\rangle}{\sqrt{w_i}} \:.
  \label{eq | non-norm states to pseudostate transform}
\end{equation}

In the \wpcd{} method we discretise the continuum of free states. A
wave packet is defined as the energy integral over some energy
``bin'' of width $\Delta E$ with Hamiltonian eigenstates
$|\psi(E)\rangle$ with positive energy $E$ as the integrand,
\begin{equation}
  |\psi(E,\Delta E)\rangle \equiv\int_{E}^{E+\Delta E} \diff E'\:|\psi(E')\rangle \:.
  \label{eq | definition eigendifferential}
\end{equation}
It follows that $\lim_{\Delta E\rightarrow0}|\psi(E,\Delta E)\rangle =
|\psi(E)\rangle$. We can define an orthogonal wave-packet basis by
letting the bin boundaries $E$ and widths $\Delta E$ lie on a mesh
such that the bins do not overlap: $\{\mathcal{D}_i\:|\:\mathcal{D}_i\cap\mathcal{D}_j=\emptyset\:\forall\:i\neq j\}_{i=1}^{N_{\text{WP}}}$,
where $\mathcal{D}_i \equiv [E_i,E_i+\Delta E_i]$ defines the integral boundaries.
Note that $E_{i+1}=E_i+\Delta E_i$. It is straightforward to normalise this basis
and show that the wave packets have eigenenergies $e_i=E_i+\frac{1}{2}\Delta E_i$.

The expectation value of an operator $\hat{O}$ in some energy bin can
now be approximately represented in a wave-packet basis
\begin{equation}
  \langle \Psi(E)|\hat{O}|\psi(E')\rangle \approx \frac{\langle \Psi(E)|\hat{O}|\psi(E_i,\Delta E_i)\rangle}{\sqrt{\Delta E_i}} \:,
  \label{eq | non-norm states to eigendifferential transformation}
\end{equation}
where $E'\in \mathcal{D}_i$. As expected, the quality of this
approximation is subject to the bin widths $\Delta E_i$.  It can be
reasoned
\cite{Rubtsova20072025,kukulin2007wave,PhysRevC.79.064602,Kukulin2003404},
on behalf of Eqs.~\eqref{eq | non-norm states to pseudostate
  transform} and \eqref{eq | non-norm states to eigendifferential
  transformation}, that the EQ weights are approximately given by
\begin{equation}
  w_i \approx \Delta E_i \:.
  \label{eq | EQ weigth identity}
\end{equation}
In short, we have a method for approximating the
EQ weights in a wave-packet basis, with which we can express the
spectrum of a scattering operator and thereby effectively solve
scattering problems.

We now proceed to approximately represent the pseudostates as
Hamiltonian eigenstates in a wave-packet representation. To this end,
we setup a wave-packet equivalent of a partial wave by generalising
Eq.~\eqref{eq | definition eigendifferential} and thus define a normalised
free wave-packet (\fwp{}) as
\begin{equation}
  |x_i\rangle \equiv \frac{1}{\sqrt{N_i}}\int_{\mathcal{D}_i}k\diff k\: f(k)|k\rangle \:,
  \label{eq | definition free wave-packet}
\end{equation}
where $f(k)$ is a weighting function and $N_i$ is a normalisation
constant. We obtain energy or momentum wave-packets
using weighting functions $f(k)=1$ or
$f(k)=\sqrt\frac{k}{\mu}$, respectively, where $\mu=\frac{m_N}{2}$ is the
reduced mass. As above, these two types of wave
packets have eigenvalues
\begin{align}
	\hat{h}_0|x_i\rangle &= \left(E_i+\frac{1}{2}\Delta E_i\right)|x_i\rangle \:,
	\label{eq | free energy WP eigenvalue} \\
	\hat{p}|x_i\rangle &= \left(k_i+\frac{1}{2}\Delta k_i\right)|x_i\rangle \:,
	\label{eq | free momentum WP eigenvalue}
\end{align}
where $\hat{p}$ is the momentum operator.
The normalisation of the momentum and energy wave-packets is given by
the bin widths, i.e. $N_i = \Delta k_i$ or $N_i = \Delta E_i$,
respectively. Operators represented in a basis of partial-wave
plane-wave states are related to the \fwp{} representation via
\begin{equation}
  \langle q'|\hat{O}|q\rangle \approx \frac{f(q)f(q')}{\sqrt{N_iN_j}}\frac{1}{q'q}\langle x_i|\hat{O}|x_j\rangle \:,
  \label{eq | operator in cont and disc representation}
\end{equation}
where $q'\in \mathcal{D}_i,\:q\in\mathcal{D}_j$. In this work we have
implemented energy as well as momentum wave-packets, and have found no
significant difference or advantage of either choice.

\begin{figure}
  \centering
  \includegraphics[width=0.9\columnwidth]{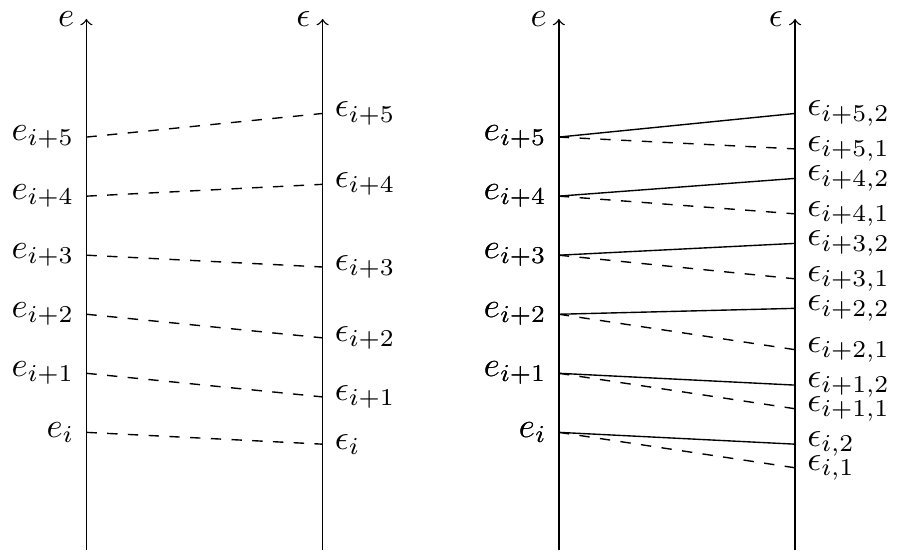}
  \caption{Left: Shift in \fwp{} energies $e_i\equiv
    E_i+\frac{1}{2}\Delta E_i$ to \swp{} energies
    $\epsilon_i$. Right: Shift and splitting of degenerate \fwp{} energies $e_i$
    into energies $\epsilon_{i,1}$ and
    $\epsilon_{i,2}$ shown by solid and dashed lines,
    respectively. Note that a solid and a dashed line from two
    different energies $e_i$ and $e_j$ will not cross
    \cite{RUBTSOVA2015613},\cite{TheorMathPhys145.1711.1726}.}
  \label{fig | wpcd splitting}
\end{figure}

The eigenstate wave-packets $|z_i\rangle$ of the full Hamiltonian with
positive energy, or scattering wave packets (\swp{}s), are given by
the eigenvalue equation $\hat{h}|z_i\rangle=\epsilon_i|z_i\rangle$ for
energies $\epsilon_i>0$. Similarly, the bound-state\footnote{Here we
  change bound-state notation from Eq.~\eqref{eq | oeprator in
    discrete basis} such that $|\psi_k^b\rangle\rightarrow
  |z_k^b\rangle$. This seems natural since $|z_k^b\rangle$ is
  expressed in terms of \fwp{}s from the diagonalisation.}
wave-packets $|z_i^b\rangle$ have energies $\epsilon_i^b<0$. We
obtain them straightforwardly via diagonalisation in a finite basis
$\{|x_i\rangle\}_{i=1}^{N_{\text{WP}}}$ of \fwp{}s, where
$N_{\text{WP}}=n+n_b$.  This operation is the most time-consuming part
in the \wpcd{} method to solve the \ls{} equation. It provides us with
a matrix of transformation coefficients $C_{ij}\equiv\langle
x_i|z_j\rangle$ such that
\begin{equation}
  H = CDC^{T} \:,
  \label{eq | eigendecomposition Hamiltonian}
\end{equation}
where $D$ is a diagonal matrix of energies $\epsilon_i^b$ and $\epsilon_i$, and
$H_{ij}=\langle x_i|\hat{h}|x_j\rangle$. In order to solve the \ls{}
equation on the form of Eq.~\eqref{eq | LS equation with full
  resolvent}, we must define a wave-packet representation of the
outbound scattering states. To define \swp{}s,
we must construct the bin boundaries
$[\mathcal{E}_i,\mathcal{E}_{i}+\Delta \mathcal{E}_{i}]=[\mathcal{E}_i,\mathcal{E}_{i+1}]$ of $|z_i\rangle$ such that the
positive eigenenergies are given by
\begin{equation}
  \epsilon_i = \mathcal{E}_i + \frac{1}{2}\Delta \mathcal{E}_{i} \:.
  \label{eq | z_i eigenenergies}
\end{equation}
This is no different than the wave-packet eigenvalues in Eqs. \eqref{eq
  | free energy WP eigenvalue} and \eqref{eq | free momentum WP
  eigenvalue}. This shift of the \fwp{} energies are indicated in the left panel in Fig.~\ref{fig | wpcd splitting}. However, it is not possible to construct the
bin-boundaries $\mathcal{E}_i$ exactly since Eq.~\eqref{eq | z_i
  eigenenergies} only provides $n$ equations while there are $n+1$
boundaries. Therefore, the following scheme \cite{PhysRevC.94.024328}
can be used to approximate the bin boundaries of $|z_i\rangle$,
\begin{align}
  \begin{split}
    \mathcal{E}_1 &\equiv 0 \:,\\ 
    \mathcal{E}_i &\equiv \frac{1}{2}\left(\epsilon_{i-1} + \epsilon_{i}\right) \:,\\
    \mathcal{E}_{n+1} &\equiv \epsilon_n + \frac{1}{2}\left(\mathcal{E}_{n} - \mathcal{E}_{n-1}\right) \:,
  \end{split}
  \label{eq | pseudostate wave packet boundaries}
\end{align}
such that they yield approximate eigenvalues $\bar{\epsilon}_i$,
\begin{equation}
  \bar{\epsilon}_i \equiv \mathcal{E}_{i} + \frac{1}{2}\Delta\mathcal{E}_{i} \approx \epsilon_i \:.
\end{equation}

In the case of $K$ coupled channels, the \fwp{} energies are
degenerate and will be split\footnote{Naturally, $K\leq2$ for \NN{}
  scattering.} in the full Hamiltonian eigendecomposition. A \fwp{}
with energy $e_i = E_i+\frac{1}{2}\Delta E_i$ will give rise to $K$
\swp{}s $|z_{i,\varkappa}\rangle$ with energies
$\epsilon_{i,\varkappa}$, corresponding to each coupled state
$\varkappa=1,2,\ldots,K$. The energies are typically ordered such that
$\epsilon_{i,\varkappa}<\epsilon_{i,\varkappa+1}\:\forall\:i,\varkappa$\cite{PhysRevC.81.064003}. Furthermore,
this prevents mixing between levels with different energies
\cite{RUBTSOVA2015613},
i.e. $\epsilon_{i,\varkappa}<\epsilon_{j,\varkappa+1}\:\forall\:i<j$. Therefore,
we use the boundary construction scheme in Eq.~\eqref{eq | pseudostate
  wave packet boundaries} such that $|z_{i,\varkappa}\rangle$ have
boundaries given by the energies $\epsilon_{i,\varkappa}$. Note it is
very important to construct the Hamiltonian matrix with degenerate
\fwp{} bases representing each coupled state, i.e. $|x_{i,1}\rangle =
|x_{i,2}\rangle$, see the right panel of Fig.~\ref{fig | wpcd
  splitting}.

The approach presented so far works fine for short-range \NN{}
potentials. Likewise, the eigenspectrum of the long-range Coulomb
Hamiltonian can be straightforwardly described using wave packets, but
these must then be constructed from Coulomb wave functions instead of
free states $|q\rangle$ \cite{TheorMathPhys145.1711.1726} as in
Eq.~\eqref{eq | definition free wave-packet}. In \wpcd{}, we
automatically ``smooth'' out the typical low-momentum singularities
presented by the Coulomb Hamiltonian. This means that the formalism of
\wpcd{} works well for both the short- and long-range
parts of the interaction, but it is necessary to treat them separately. In this
work we have not studied Coulomb wave packets as we only
consider neutron-proton scattering.

\subsection{\wpcd-method for solving the Lippmann-Schwinger equation\label{sec:WPCD}}
Following Eq.~\eqref{eq | operator in cont and disc representation}, we
can relate elements of the $T$-matrix in a continuous partial-wave basis and a
\fwp{} basis via
\begin{equation}
  T_{l'l}^{sJ}(q',q;E) = \langle q'|\hat{T}_{l'l}^{sJ}(E)|q\rangle \approx \frac{f(q)f(q')\langle x_i|\hat{T}_{l'l}^{sJ}(E)|x_j\rangle}{\sqrt{N_iN_j}qq'} \:.
\end{equation}
If we use the full resolvent $\hat{g}(E)$, defined in connection with
Eq.~\eqref{eq | LS equation with full resolvent}, we can write
\begin{equation}
  \hat{T}(E) = \hat{v} + \hat{v}\hat{g}(E)\hat{v} \:.
\end{equation}
such that in a partial-wave-projected wave-packet basis we obtain
\begin{align}
  \begin{split}
    \langle x_i|\hat{T}_{l'l}^{sJ}(E)|x_j\rangle = &\langle x_i|\hat{v}_{l'l}^{sJ}|x_j\rangle\\
    &+ \sum_{l''}\sum_{k}^{n_b}\frac{\langle x_i|\hat{v}_{l'l''}^{sJ}|z_k^b\rangle\langle z_k^b|\hat{v}_{l''l}^{sJ}|x_j\rangle}{E-\epsilon_k^b}\\
    &+\sum_{l''}\sum_{k=1}^{n}\langle x_i|\hat{v}_{l'l''}^{sJ}|z_k\rangle\\
    &\times\langle z_k|\hat{g}(E)|z_k\rangle\langle z_k|\hat{v}_{l''l}^{sJ}|x_j\rangle \:.
    \label{eq | LS equation in WP basis}
  \end{split}
\end{align}
Note that for a realistic potential we should only have $n_b=1$ (the
deuteron). The full resolvent is given by the full Hamiltonian, of
which $|z_k\rangle$ are eigenstates. In such a basis we can derive a
closed form expression for $\hat{g}(E)$, see~\ref{app |
  resolvent calculation}. In the \wpcd{} representation of
Eq.~\eqref{eq | LS equation in WP basis} all energy-dependence is
straightforwardly evaluated via the resolvent. We can therefore find
an on-shell $T$-matrix element via simple summation of the scattering
wave-packets. This is an important advantage of using the \wpcd{}
method for simulating scattering processes.
	
Note, however, that the resolvent has a logarithmic singularity for
$E=\mathcal{E}_k$ or $E=\mathcal{E}_{k+1}$. We handle this by
averaging with respect to the on-shell energy $E$,
\begin{equation}
  g_i^k \equiv \frac{1}{\Delta\mathcal{E}_k}\int_{\mathcal{D}_k}\langle z_i|g(E)|z_i\rangle\diff E\:.
  \label{eq | resolvent energy averaging}
\end{equation}

The nuclear potential, represented in a free wave-packet basis as
\begin{align}
  \begin{split}
    \langle x_i|\hat{v}_{l'l}^{sJ}|x_j\rangle = \frac{1}{\sqrt{N_iN_j}}\int_{\mathcal{D}_i}\int_{\mathcal{D}_j} k'k\diff k'\diff k \: &f(k')f(k)\\ &\times\langle k'|\hat{v}_{l'l}^{sJ}|k\rangle \:,
  \end{split}
  \label{eq | WP potential matrix identity}
\end{align}
usually vary mildly across a typical momentum-bin $\mathcal{D}_i$. It
is therefore often sufficient to use a midpoint approximation to
evaluate the integral. This offers a significant reduction in
computational cost. In a momentum wave-packet
basis, the midpoint approximation is simply given by
\begin{equation}
  \langle x_i|\hat{v}_{l'l}^{sJ}|x_j\rangle \approx \bar{k}_i\bar{k}_j\sqrt{\Delta k_i \Delta k_j}\langle \bar{k}_i|\hat{v}_{l'l}^{sJ}|\bar{k}_j\rangle \:,
\end{equation}
where $\bar{k}_i =
\frac{k_i + k_{i+1}}{2}$ are the bin midpoints.\\

Here we summarise the necessary steps to implement the \wpcd{} method
for \NN-scattering calculations.
\begin{enumerate}
\item Distribute the bin boundaries for the free wave-packets and
  choose a weighting function $f(k)$, see Eq.~\eqref{eq | definition
    free wave-packet}. Recommended options are e.g. uniform
  (equidistant), Gauss-Legendre, or Chebyshev distributions. The
  results in this work are based on a Chebyshev distribution. The
  Chebyshev distribution for $N_{\text{WP}}$ points, $\{y_j\}_{j=1}^{N_{\text{WP}}}$, is defined
  by \cite{RUBTSOVA2015613}
  \begin{equation}
    \hspace{1cm} y_j = \alpha\tan^t\left(\frac{2j-1}{4N_{\text{WP}}}\pi\right) \:, \: j=1,\ldots,N_{\text{WP}} \:,
    \label{eq | chebyshev distribution}
  \end{equation}
  where $t$ is a ``sparseness degree'' and $\alpha$ is a scaling parameter. Let $y_k$ be either the
  momentum or energy bin boundaries, and we then set the initial boundary $y_0=0$.
  For our simulations we have used
  momentum wave packets ($f(k)=1$), $t=2$, and $\alpha=100$ MeV.
  
\item Diagonalise the Hamiltonian in a free
  wave-packet basis. This yields a set of energy eigenvalues
  $\epsilon_i$ and accompanying matrix of eigenvectors $C$ as
  according to Eq.~\eqref{eq | eigendecomposition Hamiltonian}.
  
\item Construct bin-boundaries $\mathcal{E}_i$ for the pseudostate wave-packets using
  the set of energy eigenvalues and Eq.~\eqref{eq | pseudostate wave
    packet boundaries}.
  
\item Express the full resolvent $g(E)$ in the scattering wave-packet
  basis, see~\ref{app | resolvent calculation}. In this work
  we use the energy-averaged resolvent from Eq.~\eqref{eq | resolvent energy averaging}
  to avoid singularities.
  
\item Obtain approximate $T$-matrix elements in the free wave-packet basis
  by evaluating the \ls{}-equation
  \begin{equation}
    T_{ij}(E) = V_{ij} + (VC)_{ik}g_{kk}(E)(VC)_{jk} \:,
    \label{eq | on-shell T-mat element in fWP}
  \end{equation}
  where $(VC)_{ik} \equiv V_{ij}C_{jk}$, $V_{ij}\equiv\langle
  x_i|\hat{v}|x_j\rangle$, $g_{kk}\equiv\langle
  z_k|\hat{g}(E)|z_k\rangle$, and $C$ is given by Eq.~\eqref{eq |
    eigendecomposition Hamiltonian}.
\end{enumerate}

The $T$-matrix can be transformed to a continuous plane-wave basis
using Eq.~\eqref{eq | operator in cont and disc
  representation}. However, it is important to note that with energy
averaging there is ambiguity in which value $q',q\in\mathcal{D}_i$ to
 choose, as the wave-packet representation gives the same value for all choices.
 While we discuss this further in Section~\ref{sec | phase shifts}, we find it
more beneficial to choose $q',q$ such that they match the wave-packet eigenvalues,
given in Eqs.~\eqref{eq | free energy WP eigenvalue}-\eqref{eq | free momentum WP eigenvalue}.

\section{Numerical complexities of the \mi{} and \wpcd{} methods}
\label{sec | numerical complexities}
Here we present the minimum number
of floating-point operations (FLOP) required for computing
on-shell $\langle q|T_{l'l}^{sJ}|q\rangle$ matrix elements at $n_E$ different
values of the scattering energy. We focus on the \mi{} and \wpcd{}
methods, presented in Secs.~\ref{sec:MI} and~\ref{sec:WPCD},
respectively. We also assume that the potential matrix $\langle p'
|V_{l'l}^{sJ}|p\rangle$ is pre-computed and available in memory at
the outset. Typically, in both methods we use matrices of sizes
$n\times n$ for $n<100$. To avoid confusion, we let $N_Q$ denote
the number of quadrature points in the case of the \mi{} method, while
$N_{\text{WP}}$ denotes the number of wave packets
in the case of the \wpcd{} method. We retain $n$ to symbolise a basis size
in general, regardless of method.

\subsection{MI complexity}
\label{sec | GQ complexity}
Given a quadrature grid with $N_{\text{Q}}$ points, the MI method
first requires setting up the wave matrix~\eqref{eq | definition
  wave matrix} at each scattering energy. Naively, the complexity of
the matrix construction is dominated by the matrix-matrix product of
the potential $V$ with the resolvent $g_0$. Since the resolvent is
diagonal, it is more efficient to multiply each row of $V$ with the
diagonal element of $g_0$, yielding $n+1$ scalar-vector
multiplications in a single $F$-matrix construction with numerical
complexity according to
\begin{equation}
  O(F) = 2(N_{\text{Q}}+1)^2 + 2(N_{\text{Q}}+1)\:,
\end{equation}
where the factor of 2 is due to $g_0$ being complex.\\

The on-shell $T$-matrix element is obtained from the last element of
the $T$-matrix, i.e.
\begin{equation}
  T(q,q;E) = T_{N_{\text{Q}}+1,N_{\text{Q}}+1} = \sum_{i=0}^{N_{\text{Q}}+1}F^{-1}_{N_{\text{Q}}+1,i}V_{i,N_{\text{Q}}+1}\:,
\end{equation}
where we have expressed it using matrix inversion. Matrix inversion
typically requires $O(n^3)$ operations for an $n\times n$
matrix. However, solving the \ls{} linear system,
\begin{equation}
  FT = V \:,
\end{equation} 
is more advisable, and can be done very efficiently in two steps:
Firstly, we perform a lower-upper (LU) decomposition, where a square
matrix $A$ is expressed as the product of a lower-triangular matrix
$L$ with an upper-triangular matrix $U$,
\begin{equation}
  A = LU\:,
\end{equation}
which requires $O\left(\frac{2}{3}n^3\right)$
operations. The decomposition allows for $AX=B$ to be solved for each
column of $X$ using $O(2n^2)$ operations. For complex
matrices these numbers increase by a factor 4 for both routines.\\

In summary, the \mi-method has a total computational complexity of (note the
linear scaling with the number of on-shell energies $n_E$)
\begin{align}
  O_{\text{MI}}(T) &= 2n_E\left(\frac{4}{3}(N_{\text{Q}}+1)^3 + 5(N_{\text{Q}}+1)^2 + N_{\text{Q}}+1\right) \:,\label{eq | GQ complexity model}\\
  O_{\text{MI}}(K) &= n_E\left(\frac{2}{3}(N_{\text{Q}}+1)^3 + 5(N_{\text{Q}}+1)^2 + N_{\text{Q}}+1\right) \:,
\end{align}
where $O_{\text{MI}}(K)$ shows the cost of using a
$K$-matrix approach instead. The difference is simply replacing the
cost of doing complex calculations with real calculations. We see
there is roughly a factor 4 speedup in using a $K$-matrix
representation instead of a $T$-matrix representation.

\subsection{\wpcd{} complexity}
\label{sec | WPCD complexity}
The complexity of the \wpcd-method is dominated by three kinds of
linear algebra tasks i) the Hamiltonian matrix
diagonalisation, ii) two matrix-matrix products, and iii) one matrix addition
(see steps 1-5 in Sec.~\ref{sec:WPCD}). We let $N_{\text{WP}}$ be the
number of wave packets and $n_E$ the number of scattering energies
as before. By ``parallelism'' we refer to the fact that \wpcd{} can
solve the scattering problem for several energies at once, following a single
Hamiltonian diagonalisation.

For this reason we have fully implemented the \wpcd{} method on a GPU
utilising the CUDA interface with the cuBLAS \cite{cublas} and
cuSOLVER \cite{cusolver} libraries for linear operations. A more
detailed outline of the GPU code is provided in~\ref{app |
  gpu code}. Sequential algorithms are usually easier to handle, and
not all hardware allow for optimal parallelisation due to, for
example, slow computer memory transfer. Thus, we present complexity
models for both the sequential and parallel energy-evaluations of the
\wpcd{} method. Note, however, that this is simply a comparison of
FLOP models that do not account for memory transfer times or detailed
processor architecture. We emphasise that all \wpcd{} results
presented in this paper were calculated using our parallel GPU-code,
and that the sequential complexity is presented purely to provide
further insight.

The \mi{} method can also be implemented efficiently on a GPU. This
approach will not exhibit the same inherent parallelism with respect
to multiple on-shell calculations. However, there does exist efficient
parallel routines for solving linear systems on the form of
Eq.~\eqref{eq | LS equation in matrix form with wave matrix}.

\subsubsection{Sequential complexity}
The sequential mode for the \wpcd{} method means solving the \ls{} equation
for each on-shell energy in sequence rather than simultaneously. This
suggests using a CPU, rather than a GPU, since CPUs typically have a
faster clock frequency and can thus iterate through the on-shell
energies faster than a GPU. Of course, CPUs today are multicore and
with several processing units, but note that they might not have a
number of cores equal to or greater than $n_E$. The analysis below
assumes a scenario where a processing unit is working with a single
computational core.

\begin{enumerate}
\item A real-valued \NN{} Hamiltonian is efficiently diagonalised using
  QR factorisation or a divide-and-conquer algorithm, both of which
  are usually used together with Householder transformations. The
  divide-and-conquer algorithm has a complexity of
  $O\left(\frac{8}{3}n^3\right)$ for getting both
  eigenvectors and eigenvalues of an $n\times n$ matrix.
  
\item Next, we evaluate the matrix-matrix product of the potential
  matrix $V$ and the coefficient matrix $C$ in the rightmost term in
  Eq.~\eqref{eq | on-shell T-mat element in fWP}. Square
  matrix-matrix multiplication requires $2n^3-n^2$ FLOP in a
  straightforward, sequential approach.
  
\item The LS equation for $T_{ii}(E\in\mathcal{D}_i)$ is evaluated via
  simple summation and multiplication. The sum,
  \begin{equation}
    \sum_{j=1}^{N_{\text{WP}}} (VC)_{ij}g_{jj}(E\in\mathcal{D}_i)(VC)_{ji} \:,
    \label{eq | LS WP sum}
  \end{equation}
  involves two multiplications per term, and the sum will require
  $N_{\text{WP}}-1$ additions. There is also the addition of the first
  term in Eq.~\eqref{eq | on-shell T-mat element in fWP}. These
  operations must be done for every on-shell energy, resulting in a
  complexity of $6N_{\text{WP}}\times n_E$ FLOP. Note that if we use
  energy averaging we only require a single evaluation for each
  on-shell wave packet, giving an upper limit $n_E\leq N_{\text{WP}}$.
\end{enumerate}
The \wpcd{} sequential complexity is then given by
\begin{equation}
  O_{\text{WPCD,seq.}}(T) = 6N_{\text{WP}}^3 - N_{\text{WP}}^2 + 6N_{\text{WP}}\times n_E \:,
  \label{eq | WPCD serial flop model}
\end{equation}

To summarise, the overall complexity of the \mi{} method scales
linearly with the number of scattering energies $n_E$, see
Eq.~\eqref{eq | GQ complexity model}. For the \wpcd{} method we get
all the on-shell energy scattering-solutions from a single Hamiltonian
diagonalisation, giving a very ``cheap'' scaling with $n_E$.

\subsubsection{Parallel complexity}	
The parallel \wpcd{} approach is based on simultaneous solutions of
the \ls{} equation for all on-shell energies. Contrary to the
sequential approach, we assume a sufficiently large set of processors
(like in a GPU) to handle all on-shell energies at once. This will
remove the factor $n_E$ in the complexity model~\eqref{eq | WPCD
  serial flop model} such that the cubically-scaling term will clearly
become the dominating one. It can be combatted by making use of
hardware-specific, massively parallel algorithms for matrix
diagonalisation and matrix-matrix multiplications. Here
we present the parallel approach to the same steps as above, and in
the same ordering. We assume we have $p$ compute processors, or
threads, available.
\begin{enumerate}
\item The parallel cyclic-order Jacobi method is a parallel approach
  to the Jacobi eigenvalue algorithm: a method based on finding a
  similarity transformation of a matrix to its diagonal form by
  repeated Jacobi rotations (see e.g. Ref.~\cite{GoluVanl96}). This
  method has held major appeal for parallel computing due to its
  inherent parallelism compared to other eigenvalue-finding
  routines. However, parallel optimisation of the basic algorithm
  depends greatly on how a set of processors is organised with regards
  to communication and memory.  Therefore, there is no general
  complexity model for the method---any algorithm should be written
  with a specific hardware in mind. One approach can be found in
  Ref.~\cite{Pourzandi1994} with a demonstrated complexity
  $O\left(\frac{n^3}{p}\log n\right)$ for convergence for a
  symmetric and real $n\times n$ matrix.
  
\item Parallel matrix-matrix multiplication algorithms is an ongoing
  field of research (see e.g. Ref \cite{ABDELFATTAH2020188}). While
  there exist very efficient methods, such as Cannon's
  algorithm~\cite{BAE20142230}, they depend highly on hardware and
  matrix characteristics (sparsity, symmetry, etc). The
  divide-and-conquer approach is a very general and straightforward
  way to parallelise matrix multiplications. In essence, we divide the
  matrices into $M=\ceil*{\frac{n^2}{p}}$ submatrices of sizes
  $m\times m=p$. If combined with on-chip shared memory between the
  processors, this permits all processors to simultaneously work on
  calculating the product while minimising processor-to-processor
  communication time. The method is explained excellently in
  literature such as \cite{GoluVanl96}. This simple method has a
  theoretical minimum complexity of
  $O\left(n\times M\right)$ when performed in
  parallel\footnote{This is not the full picture. Often, a set of
    processors is divided in groups of processors with limited shared
    memory that cannot fit all $p$ elements from each matrix in the
    matrix-matrix product, and one has to define more submatrices
    than given by $M$. This introduces yet another complication to the
    parallel complexity model that has to do with shared memory size,
    which we will not account for here.}.
  
\item There is limited parallel optimisation to be gained in
  Eq.~\eqref{eq | LS WP sum}. We can multiply each term of the
  summation in parallel, and do the whole summation sequentially for
  the sake of simplicity. The result is a complexity of
  $O(2(N_{\text{WP}}+2))$ FLOP. Note, however, that the
  inherent parallelism of the \wpcd{} method allows us to do the
  summation simultaneously for all $E$ and thereby effectively
  removing the factor $n_E$ seen so far in the complexity models.
\end{enumerate}

In conclusion, assuming $p>N_{\mathrm{WP}}$ and adequately large
shared memory (see point 2 above), a somewhat optimal and parallel
\wpcd{} complexity model is given by
\begin{align}
	\begin{split}
	O_{\text{WPCD,par.}}(T) = &\ceil*{\frac{N_{\text{WP}}^3}{p}}\log(N_{\text{WP}}) \\&+ N_{\text{WP}}\ceil*{\frac{N_{\text{WP}}^2}{p}} + 2(N_{\text{WP}}+2) \:.
	\end{split}
  \label{eq | WPCD parallel flop model}
\end{align}

\begin{figure}
	\centering
	\includegraphics[width=0.7\columnwidth]{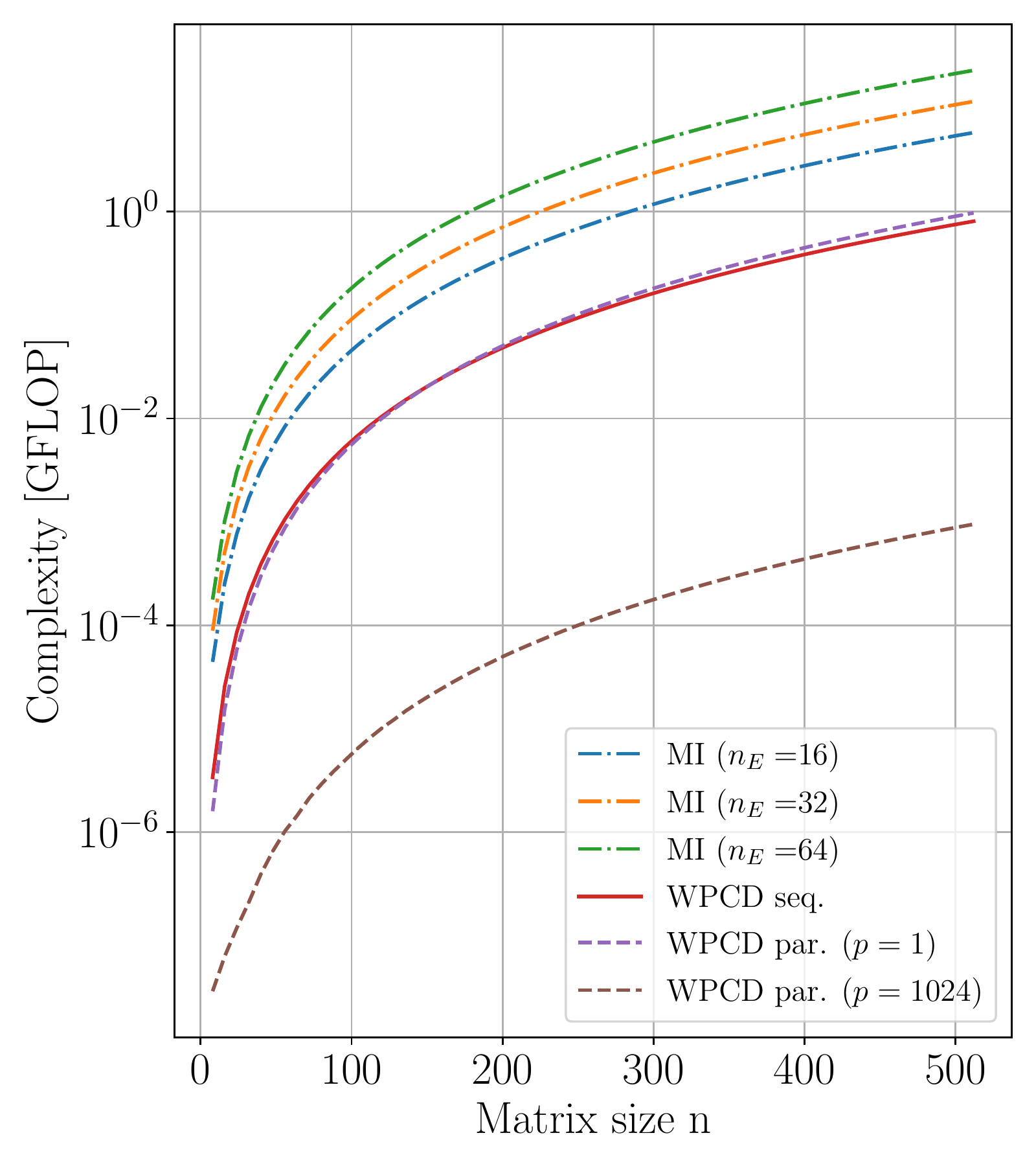}
	\caption{Complexity, as measured in FLOP, for the \mi{} and
          \wpcd{} methods as a function of $n$, where $n$ represents
          the number of quadrature points ($N_{\text{Q}}$) for \mi{}
          and the number of wave packets ($N_{\text{WP}}$) for
          \wpcd{}. Furthermore, $n_E$ is the number of on-shell $T$-matrix
          evaluations (note $n_E=n=N_{\text{WP}}$ for \wpcd{}).}
	\label{fig | theoretical complexities}
\end{figure}

We see from Fig.~\ref{fig | theoretical complexities} that the
efficiency of the parallel \wpcd{} approach scales very well with the
number of available processors. We also see that for a single
processor ($p=1$) the parallel and sequential approaches are roughly
equal (while not taking into account any overhead in memory transfers),
while for a realistic value $p=1024$ the complexity model for the
parallel approach demonstrates a clear advantage. The value $p=1024$
corresponds to the current limit on threads with shared memory for
typical Nvidia GPUs.

\section{Continuum-discretised neutron-proton scattering computations}
\label{sec | method accuracy and precision}
In this section we present a detailed analysis of the precision and
accuracy of the \wpcd{} method for computing neutron-proton (\np{})
scattering observables and phase shifts. For all calculations, we
employed the optimised next-to-next-to-leading-order chiral potential
$\text{N2LO}_{\text{opt}}$ \cite{PhysRevLett.110.192502}. The primary
goal is to analyse the trade-off between minimum computational cost
and maximum method accuracy in the \wpcd{} method, and contrast this
to the conventional \mi{} method. Note that there is no problem in
obtaining highly accurate results from either method. We simply focus
on the performance and accuracy of the \wpcd{} method as we reduce the
number of wave-packets such that all objects fit in the fast on-chip
shared memory on the GPU. In our analysis we consider the \mi{} method
with $N_{\text{Q}}=96$ Gauss-Legendre points, see Eq.~\eqref{eq | LS
  equation in operator form}, to yield virtually exact results. For
brevity, we will refer to such numerically converged calculations as
\emph{exact}.
\begin{figure*}
  \centering
  \includegraphics[width=\columnwidth]{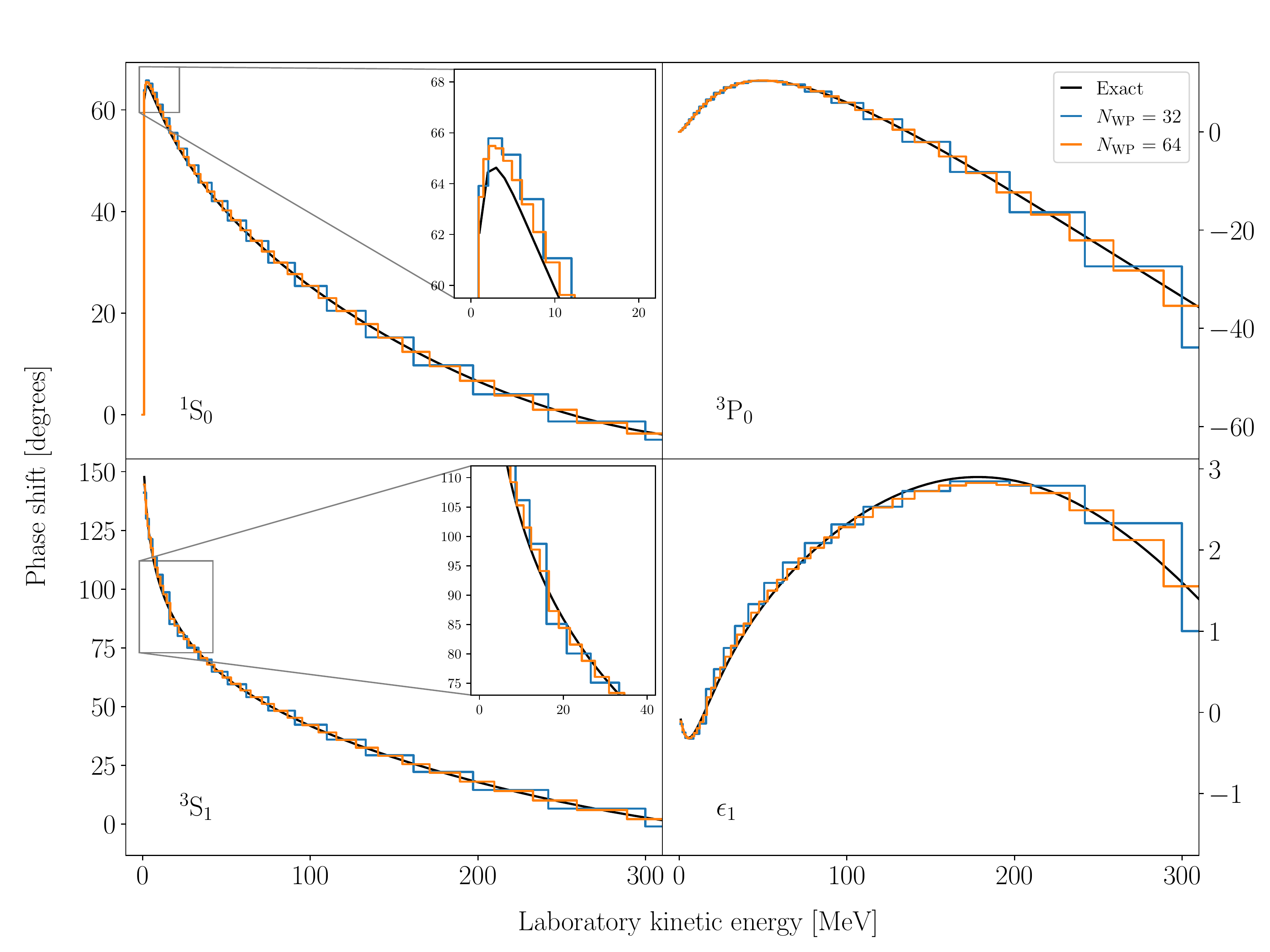}
  \caption{Phase shifts calculated using the \wpcd-method using
    \NWP=32 (blue) and \NWP=64 (orange) wave packets.  Numerically
    \emph{exact} results (black) were obtained using the \mi-method.}
  \label{fig | phases vs energy}
\end{figure*}
\subsection{Computing phase shifts \label{sec | phase shifts}}
In Fig.~\ref{fig | phases vs energy} we compare \np{} scattering phase
shifts in the $^1\text{S}_0$, $^3\text{P}_0$, and $^3\text{S}_1$
partial waves as well as the $\epsilon_{1}$ mixing angle, as obtained
from the \wpcd{} method using \NWP=32 and \NWP=64 momentum
wave-packets. The free wave-packet bin boundaries follow a Chebyshev
distribution with scaling factor $\alpha=100$ MeV and sparseness
degree $t=2$, see Eq.~\eqref{eq | chebyshev distribution}, and the
resolvent was energy-averaged according to Eq.~\eqref{eq | energy
  averaged resolvent}.

For the \wpcd{}-result,
one immediately observes a discretised, step-like version of the otherwise
smoothly varying \textit{exact} description of the scattering phase
shifts. This is entirely due to the momentum discretisation and finite
number of wave packets. Indeed, due to the energy-averaging we obtain
only one $T$-matrix element per momentum bin. In the limit
\NWP$\rightarrow \infty$, and everything else equal, we will recover
the \textit{exact} result obtained with the \mi{} method. We observe
this convergence
trend already when going from \NWP=32 to \NWP=64 wave packets. For
instance, the mixing angle moves closer to the \emph{exact} results when
increasing \NWP. Overall, the \wpcd{} method is performing as
expected, but there are two features we would like to point out:

\begin{enumerate}	
\item The values of the low-energy \SWAVE{} phase shift are
  overestimated near the peak where the phase shift turns over. This
  is likely due to the momentum-averaging of operators. The potential
  matrix in the \SWAVE{} channel ($\langle
  q'|V_{^1\textrm{S}_0}|q\rangle$) is shown in Fig.~\ref{fig | V1S0
    heatmap} for both a continuous momentum-basis and a
  $N_{\text{WP}}=32$ wave-packet basis with bins distributed according
  to Eq.~\eqref{eq | chebyshev distribution}. The potential is
  constant within each wave-packet bin as expected according to
  Eq.~\eqref{eq | WP potential matrix identity}. This makes it
  challenging to reproduce finer details of the interaction. There is
  also a discrepancy between the continuous and wave-packet values
  when the chosen $q'$-momentum is not near any bra-state bin
  midpoint. This is most distinctive in the green curves at
  $q'=205.622$ MeV, which is very close to a bin boundary.  We see in
  Eq.~\eqref{eq | WP potential matrix identity} that we average over
  momenta within two bins, and when a potential varies strongly within
  a bin such that the matrix elements at the bin boundaries are quite
  different from the bin mid-point values, this averaging is too
  coarse to mimic the potential accurately. This effect becomes less
  significant with increasing $N_{\text{WP}}$ since the grid will grow
  denser.

\begin{figure}
  \centering
  \includegraphics[width=0.7\columnwidth]{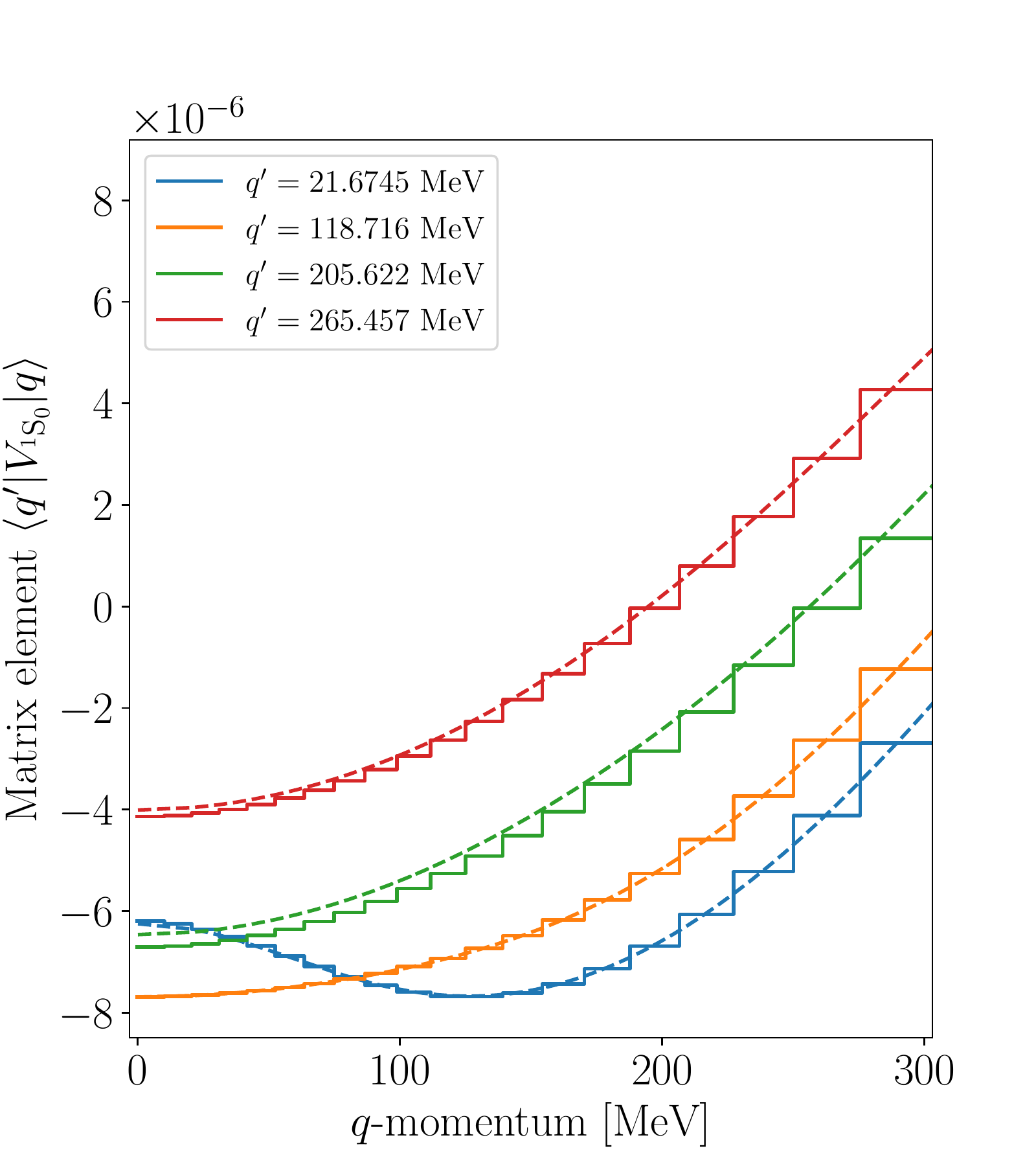}
  \caption{ The \SWAVE{} potential of NNLO$_{\rm
      opt}$~\cite{PhysRevLett.110.192502} as a function of $q$, at
    several $q'$ values. The dashed lines represent the potential
    calculated in a continuous basis, while the solid lines represent
    a wave-packet projection. The wave-packet projection was made with
    an $N_{\text{WP}}=32$ basis in a Chebyshev distribution (with
    $\alpha=100$ MeV). In the wave-packet representation, the
    bra-state bin contains $q'$. Note that the units for continuous
    representation of the potential is in units of MeV$^{-2}$ while
    the wave-packet projected potential is in units of MeV, as follows
    from the definition in Eq.~\eqref{eq | definition free
      wave-packet}.
    \label{fig | V1S0 heatmap}%
    }
\end{figure}

\item
  The \wpcd-results for \STWAVE{} show a distinctive drop at 90
  degrees.  This trend is consistent for all basis sizes and is simply
  due to the treatment of the deuteron bound state in the \wpcd{}
  framework. We extract a phase shift $\delta$ via an inverse
  trigonometric function with values for $\delta\in[-90,90]$
  degrees. A bound state is characterised by a transition
  $\delta(E+\epsilon)-\delta(E)=180$ degrees at some energy $E$ for an
  infinitesimal step $\epsilon>0$, as dictated by Levinson's theorem
  \cite{osti_4434712}. It is apparent that this transition across 90
  degrees is more difficult to reproduce for the \wpcd{} method. The
  same effect also gives rise to inaccuracies in the $\epsilon_1$
  mixing angle. However, this is of no concern when computing a
  scattering observable.
\end{enumerate}

\subsubsection{Linearly interpolating phase shifts}
\begin{figure}
  \centering
  \includegraphics[width=0.7\columnwidth]{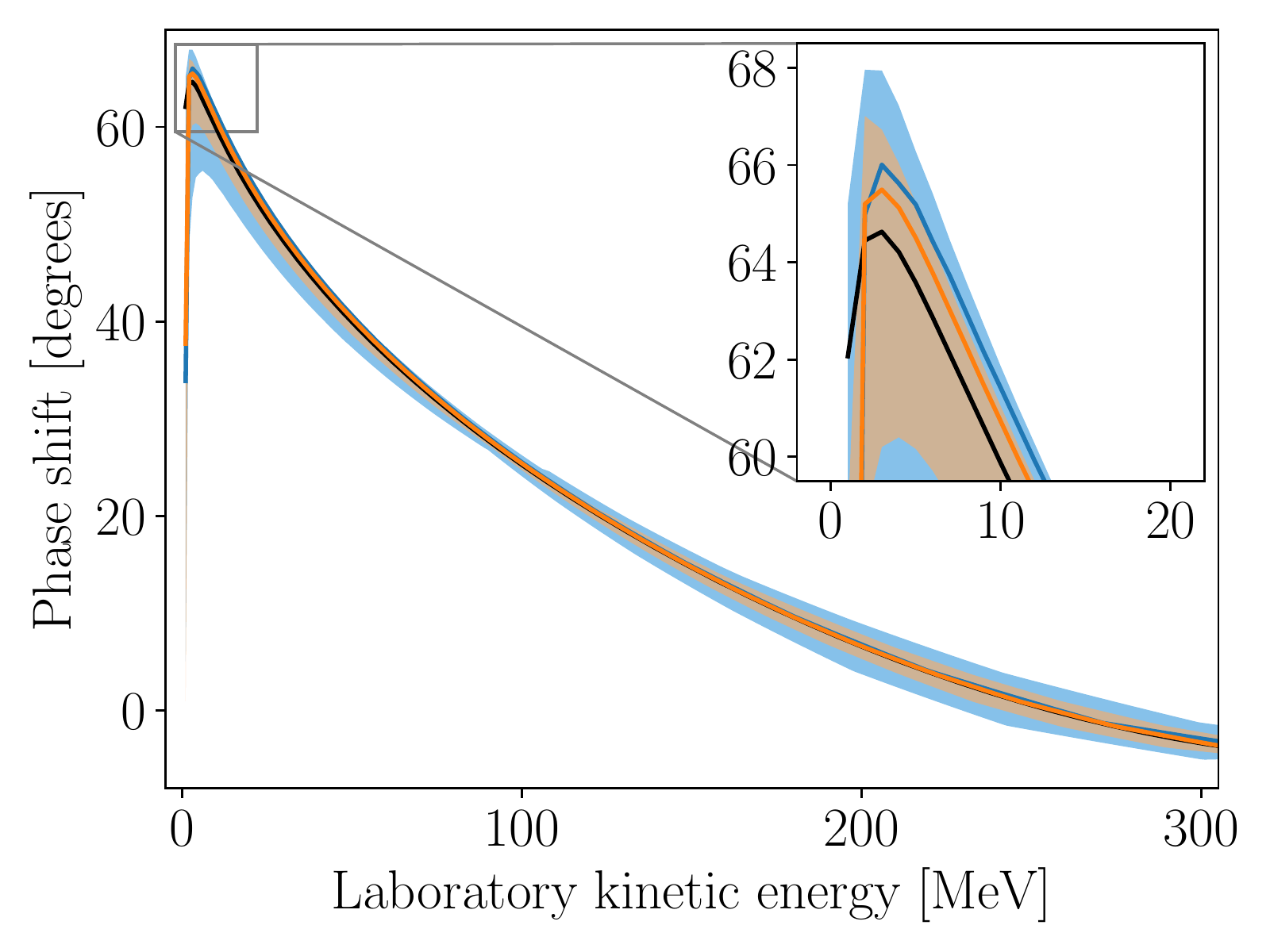}
  \caption{Phase shifts for the $^1\text{S}_0$ partial wave
    calculated using \wpcd{}. The colour coding is the same as in
    Fig.~\ref{fig | phases vs energy}, i.e. $N_{\text{WP}}=32$ (light
    blue) and $N_{\text{WP}}=64$ (light orange). The bands indicate
    the maximum discrepancy due to the choice of interpolation point
    $n$ according to Eq.~\eqref{eq | interpolation point}.}
  \label{fig |phases and SGT interpolated with variance}
\end{figure}

Although the resolution of the \wpcd{} method is limited by the
energy-averaging across each momentum-bin---i.e., yields a step-like
character of the predictions---the mid-points
$\bar{q}_i=\frac{1}{2}(q_{i-1}+q_i)$ of each bin in Fig.~\ref{fig |
  phases vs energy} are typically closest to the \emph{exact} results,
as expected from the wave-packet eigenvalues, see Eqs.~\eqref{eq |
  free energy WP eigenvalue} and \eqref{eq | free momentum WP
  eigenvalue}.  We can therefore linearly interpolate the phase shifts
$\delta_i\equiv\delta(\bar{q}_i)$ across several bins, i.e. across
scattering energies, via
\begin{equation}
  \delta(q) = \left(\frac{\delta_i-\delta_{i-1}}{\bar{q}_i-\bar{q}_{i-1}}\right)q + \left(\delta_{i-1}-\frac{\delta_i-\delta_{i-1}}{\bar{q}_i-\bar{q}_{i-1}}\bar{q}_{i-1}\right) \:.
  \label{eq | linear interpolation}
\end{equation}
for c.m. momenta $q\in[q_{i-1},q_i]$ where $q_i$ are the \fwp{} bin boundaries.
This simple and straightforward approach offers a rather precise
prediction for any on-shell scattering energy. Of course, the phase
shifts can be linearly interpolated using other points, i.e. we can
let
\begin{equation}
  \bar{q}_i = q_{i-1} + \frac{n}{m}(q_i-q_{i-1}) \:,
  \label{eq | interpolation point}
\end{equation}
for some $n\in [0,m]$, such that $n=\frac{m}{2}$ is the bin mid-point
again.
Figure~\ref{fig |phases and SGT interpolated with variance} shows the
resulting predictions using linear (mid-point) interpolation of the \SWAVE{}
phases presented in Fig.~\ref{fig | phases vs energy}. The bands
estimate the effect of varying the interpolation point. The bands span
the resulting $\delta(q)$ calculated with $m=10$ and
$n=[0.1,1,2,\ldots8,9,9.9]$ in Eq.~\eqref{eq | interpolation
  point}. As expected, mid-point interpolation yields results that are
very close to the \emph{exact} calculation. In the following we will
therefore only use this interpolation choice.

\subsection{Computing cross sections}
\begin{figure}
  \centering
  \includegraphics[width=0.7\columnwidth]{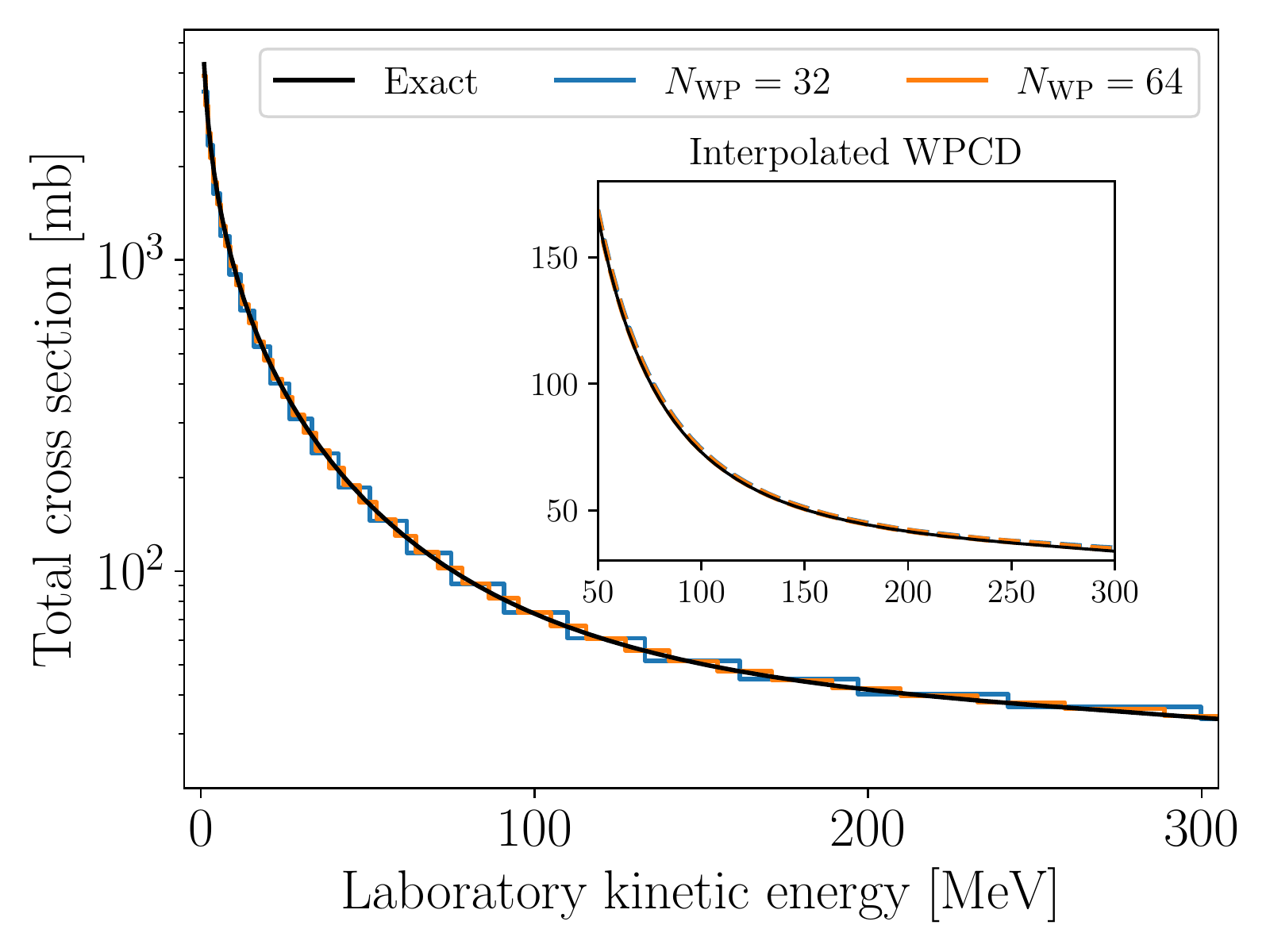}
  \caption{Total cross section calculated with the \wpcd{} method. The
    solid black line corresponds to the \textit{exact} result as
    obtained using the \mi{} method. The inset demonstrates the high
    accuracy of (mid-point) interpolated \wpcd{} results. On this scale,
    the two separate \wpcd{} predictions appear to overlap completely.}
  \label{fig | sgt binned}
\end{figure}

We compute scattering cross sections from scattering amplitudes
according to the method outlined in~\ref{app | scattering
  theory}. Figure~\ref{fig | sgt binned} shows the total cross section
obtained using the \wpcd-method using momentum-space bins with
$N_{\text{WP}}=32$ and $N_{\text{WP}}=64$. As can be seen in the
figure, the linearly-interpolated results reproduce the \textit{exact}
result rather well. On a larger scale it is nearly impossible to tell
any difference between results obtained using $N_{\text{WP}}=32$ bins
and $N_{\text{WP}}=64$ bins. To emphasise the monotonically increasing
accuracy of the \wpcd{} method as we increase the number of
momentum-space bins, we calculate the absolute value of the difference
between the \emph{exact} and the \wpcd{} prediction for the total cross
section for a range of bin resolutions, see Fig.~\ref{fig | SGT error
  heatmap}. From this it is apparent that the \wpcd{} method
converges, although slowly, for a bin partition following a Chebyshev
distribution.

\begin{figure}
	\centering
	\includegraphics[width=0.7\columnwidth]{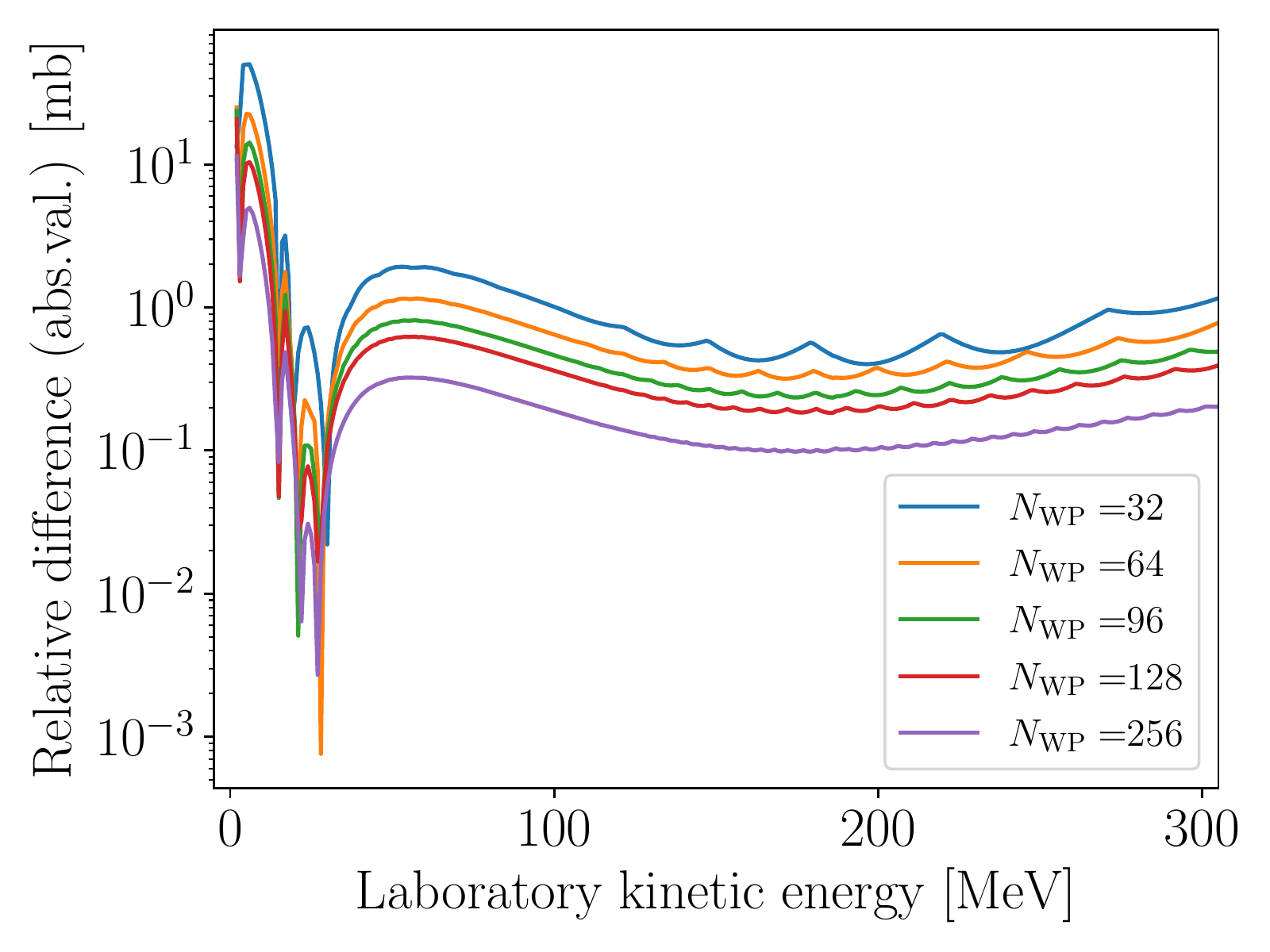}
	\caption{Absolute values of the relative difference between \emph{exact}
		results and \wpcd{} method for the total cross section, shown as a
		function of wave-packet basis size $N_{\text{WP}}$ and laboratory
		kinetic energy.}
	\label{fig | SGT error heatmap}
\end{figure}

\subsection{\wpcd{} accuracy}
There are primarily two approximations that impact the accuracy of the
\wpcd{} method: the number of bins $N_{\text{WP}}$ and the
distribution thereof. Also, the averaging of continuous states into wave
packets means we use momentum-averaged matrix representations of
operators in the LS equation. This averaging should improve with
reduced bin widths. We can easily control $N_{\text{WP}}$, and in this
section we analyse the performance of \wpcd{} with respect to
different choices of this method parameter.

To better quantify the convergence of the \wpcd{} method we study the
root-mean-square error (RMSE) with respect to the exact result for the
total cross section across a range of scattering energies. We use
the standard RMSE measure
\begin{equation}
  \text{RMSE} = \sqrt{\frac{\sum_{i=1}^{n_E}\left(\sigma_{\text{exact},i} - \sigma_{\text{method},i}\right)^2}{n_E}}\:,
\end{equation}
where $\sigma_{\text{exact},i}$ denotes the \emph{exact} results and
$\sigma_{\text{method},i}$ denotes either the \wpcd{}- or MI-calculated
total cross sections at some scattering energy $E_i$ for
$i=1,\ldots,n_E$.\\

We find that the RMSE for \wpcd{} with scattering energies
corresponding to laboratory kinetic energies $40 \leq T_{\rm lab} \leq
350$ MeV remains fairly constant at $\sim 2.0$ mb when using
$N_{\text{WP}}=16$. This is interesting for three reasons:
\begin{itemize}
\item The \wpcd{} coupled channel Hamiltonian will be of size
  $2N_{\text{WP}}\times 2N_{\text{WP}}$, i.e. we diagonalise
  $32\times32$ Hamiltonian matrices. Most GPUs today, including the
  Nvidia Tesla V100 we have used here, have 64 kB shared memory (also
  called ``on-chip'' memory) which is significantly faster to access
  than the GPU's main memory. These matrices fit entirely on the GPU
  shared memory, allowing for a strong reduction in GPU memory
  read/write demand while performing the diagonalisation.
  
\item A recent Bayesian analysis of chiral interactions suggests that
  the \cheft{} truncation error for scattering cross sections at
  next-to-next-to-leading-order is at least 2 mb, at 68\%
  degree-of-belief (DoB), and at least 5 mb at 95\% DoB
  \cite{PhysRevC.96.024003}. 

\item Regarding experimental uncertainties, the combined statistical
  and systematic uncertainties in measured np total cross sections are
  at the 1\% level~\cite{PhysRevLett.49.255, PhysRevC.63.044608}. In
  absolute terms, this amounts to uncertainties of the order $1$ mb at
  laboratory scattering energies $\gtrsim 40$ MeV. At lower energies
  the cross section increases of course, leading to slightly larger
  absolute experimental errors.
\end{itemize}

In summary, we find that the \wpcd{} method error, when using very few
wave-packets, is slightly larger than typical experimental
uncertainties but smaller than the estimated model discrepancy up to
next-to-next-to-leading-order in $\chi$EFT.

One can clearly reduce the \wpcd{} method error by increasing
$N_{\rm WP}$. However, this will also increase the computational cost
and we recommend studying the actual wall time cost in relation to,
e.g. the RMSE measure defined above. We perform such an analysis in
Sec.~\ref{sec | time profiling and program efficiency}.

Before ending this section on the method accuracy, we would like to
emphasise that the maximum total angular momentum $J_{\text{max}}$ in
the partial-wave expansion of the potential also impacts the accuracy
of the description of the scattering amplitude. For the
linearly-interpolated \wpcd{} method with $N_{\text{WP}} = 16$, where
we observe a method error of $\sim 2$ mb in the total cross section at
$T_{\rm Lab}\gtrsim 40$ MeV, and find it unnecessary to go beyond
$J_{\rm max}=6$.

\section{Method Performance}
\label{sec | time profiling and program efficiency}        
With an account of the precision and accuracy of the \wpcd{} method we
now profile its time performance. Typically, when using the \mi{}
method, we calculate every on-shell phase shift of interest by
explicitly solving the \ls{} equation, while in the \wpcd{} approach
we interpolate in-between bin mid-points, as discussed above. This
will, unsurprisingly, induce a substantial time-performance penalty in
using the \mi{} method, due to the linear scaling with the number of
energy solutions, as shown in Eq.~\eqref{eq | GQ complexity
  model}. However, we can of course linearly interpolate the phase
shifts calculated with the \mi{} method, rather than invoking an
explicit calculation at every on-shell energy. Therefore, to
facilitate a balanced comparison between the two methods, we also employed
linear interpolation to extract phase shifts when using the \mi{}
method. In our studies, this has turned out to be a highly efficient
way to speed up the calculation with the \mi{} method while
maintaining precision and accuracy of the results.

To facilitate a comparison, when using the \mi{} method we solve the
\ls{} equation at $n_E$ on-shell energies also following a Chebyshev
distribution. We then linearly interpolate the phase shifts using
these energies to calculate neutron-proton total cross sections in the $T_{\rm lab}$
energy ranges $(0,350]$ MeV and $(40,350]$ MeV for the calculation of RMSE values.

For \wpcd{} we can only vary the number of wave packets $N_{\rm WP}$
while for \mi{} we can vary both $N_{\text{Q}}$ and $n_E$, i.e. the
number quadrature points and the number of interpolation points or
on-shell energies (at bin mid-points), respectively. Figure \ref{fig |
  time vs rmse} shows the wall times for solving the \ls{} equation to
obtain cross sections, at different levels of method accuracy measured
by the RMSE value. In Tab.~\ref{tab | WPCD vs GQ time profiles} we
show a few interpretations of the figure for a handful of relevant
method parameters. As mentioned, the \wpcd{} method is implemented on
a GPU while the \mi{} results were obtained using an optimised CPU
implementation~\cite{PhysRevX.6.011019}. The time profiles were
obtained using an Nvidia Tesla V100 32 GB SMX2 and an Intel Xeon Gold
6130 for the \wpcd{} GPU and \mi{} CPU results, respectively.

\begin{figure}
  \centering
  \includegraphics[width=0.7\columnwidth]{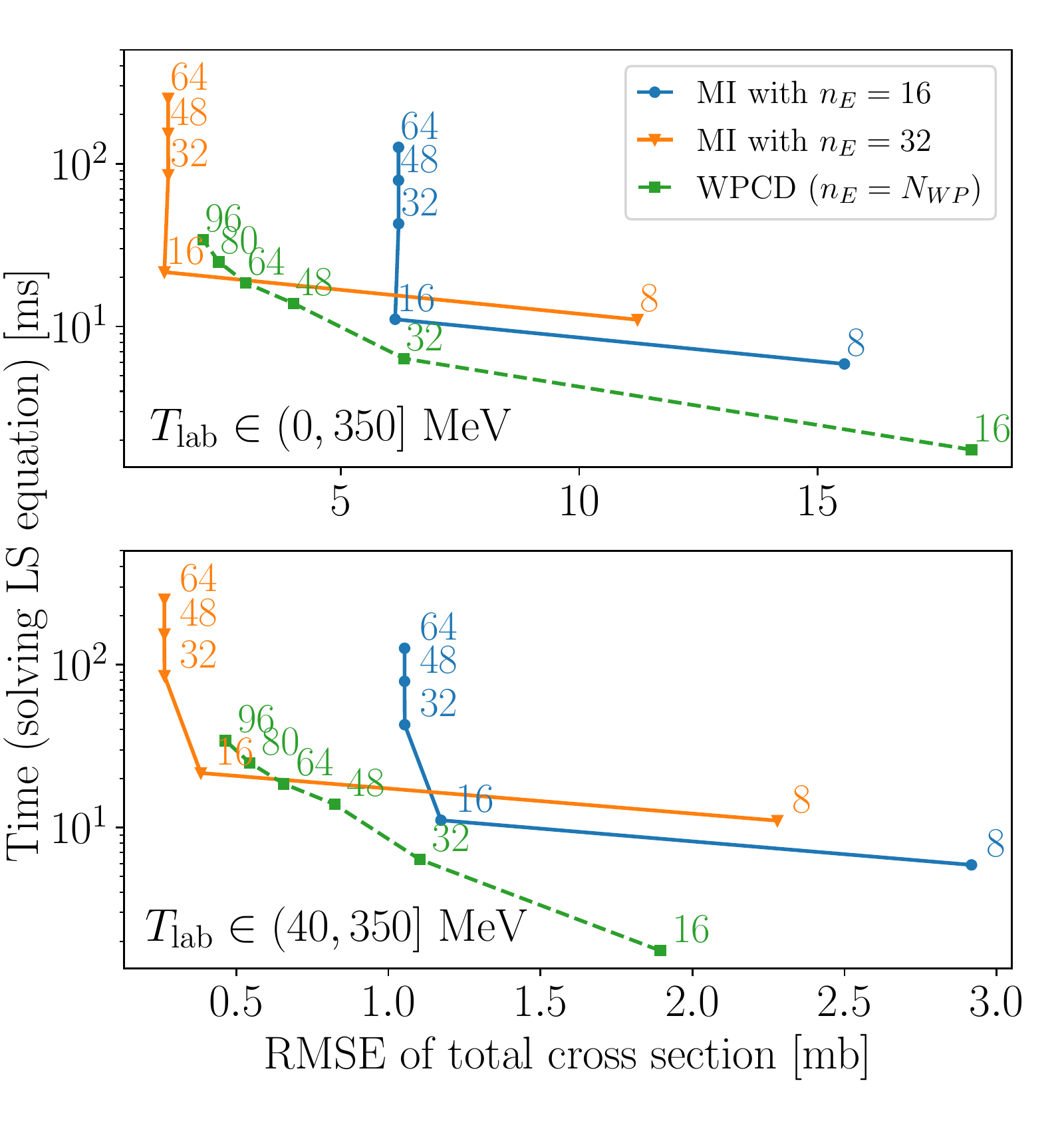}
  \caption{Measured wall times for solving the LS equation for
    different values of total cross section RMSE for
    $T_{\text{lab}}\in(0,350]$ MeV (top panel) and
  $T_{\text{lab}}\in(40,350]$ MeV (bottom panel). $n_E$ is the number
    of on-shell energies (interpolation points) at WP bin mid-points,
    according to a Chebyshev distribution. The numeric label attached
    to each data point indicates the number of Gauss-Legendre grid
    points $N_{\text{Q}}$ for the \mi{} method and the number of wave
    packets $N_{\text{WP}}$ for the \wpcd{} method. The number of
    energies used for interpolating the \mi{} results are indicated in
    the legend. See the main text for hardware specifications.
    \label{fig | time vs rmse}%
    }
\end{figure}

From our analysis we conclude that the \wpcd{} method is faster than
the \mi{} method if one can tolerate $\sim 1-5$ mb method RMSE in the
prediction of total scattering cross sections. For such applications
it would then be advisable to use $N_{\text{WP}}\gtrsim 48$ bins, on
the basis of Fig.~\ref{fig | time vs rmse}. It is worth noting that
the RMSE is dominated by contributions from scattering cross sections
below laboratory kinetic energies $T_{\rm lab} \sim 40$ MeV. Indeed,
with $N_{\rm WP}=48$ we obtain an RMSE value of $\sim4$ mb across an
interval $T_{\rm lab} \in (0,350]$ MeV. The RMSE drops to $\sim0.8$ mb
  when considering only cross sections with $T_{\rm lab}>40$
  MeV. Irrespective of how the WP bins are distributed, it is therefore
  necessary to ensure a sufficiently high density of bins below
  $T_{\text{lab}}=40$ MeV to accurately reproduce the \SWAVE{}
  phase-shift peak and the \STWAVE{} bound state. With the \wpcd{}
  method we can obtain increasingly faster solutions to the \ls{}
  equation as we reduce the number of wave-packets.
	
\begin{table}
	\centering
	\caption{Measured wall times, corresponding to the results in
          Fig.~\ref{fig | time vs rmse}, for solving the \ls{}
          equation and predicting total cross sections using the
          \wpcd{} and \mi{} methods, at different RMSE levels in the
          $T_{\text{lab}}\in(40,350]$ MeV energy region. The $n_E$
        parameter indicates the number of interpolation energies used for
        the \mi{} method.}
	\begin{tabular}{|c|c|c|c|c|l|c|}
		\hline
		\multirow{2}{*}{RMSE {[}mb{]}} & \multicolumn{3}{c|}{MI}          & \multicolumn{3}{c|}{WPCD}                       \\ \cline{2-7} 
		& $N_{\text{Q}}$ & $n_E$ & Time {[}ms{]} & \multicolumn{2}{c|}{$N_{\text{WP}}$} & Time {[}ms{]}   \\ \hline
		3.0                            & 8        & 16    & 6             & \multicolumn{2}{c|}{--}     & --            \\ \hline
		2.5                            & $10$        & 16    & 7            & \multicolumn{2}{c|}{--}     & --            \\ \hline
		2.0                            & $12$        & 16    & 8     & \multicolumn{2}{c|}{16}       & 0.5             \\ \hline
		1.5                            & $14$       & 16    & 10            & \multicolumn{2}{c|}{$24$}    & 2.5             \\ \hline
		1.0                            & $16$       & $24$    & $ 15$      & \multicolumn{2}{c|}{32}       & 6               \\ \hline
		0.5                            & 16       & 32    & 20            & \multicolumn{2}{c|}{96}       & $35$              \\ \hline
	\end{tabular}
	\label{tab | WPCD vs GQ time profiles}
\end{table}

\section{Summary and outlook}
In this study we have compared the \wpcd{} method with the standard
\mi{} method, with an emphasis on their respective efficiency,
precision, and accuracy when solving \NN{} scattering problems. This
study is done with an interest in reducing the computational costs of
Bayesian inference analyses of nuclear interaction models.

We find that the \wpcd{} method, with GPU-acceleration, is capable of
providing scattering solutions to the \NN{} \ls{} equation much faster
than a conventional \mi{} method implemented on a CPU. This is largely
due to the \wpcd{} method being capable of delivering scattering
amplitudes at several on-shell energies in an inherently parallel
fashion, and that linear interpolation across several WP bins offers
straightforward predictions for any scattering energy. However,
compared to solutions obtained using the standard \mi{} method, the
\wpcd{} method is less accurate, in particular in regions where the
scattering amplitudes vary strongly with energy. This is also expected
due to the momentum-space discretisation and approximation of the
scattering states. Nevertheless, in applications where a certain
method error can be tolerated---as for example in low-order EFT
predictions of \NN{} scattering cross sections---the GPU-implemented
\wpcd{} method presents a computational advantage. This finding makes
it particularly promising for computational statistics analyses of EFT
nuclear interactions utilising Bayesian inference methods.

We also find that the computational gain of using GPU hardware is
limited by the amount of shared memory that is available. This
constraint basically corresponds to an upper limit on the number of WP
bins that can be used for efficient computations.  We note, however,
that GPU hardware is continuously improved, and that a four-fold
increase in the size of the fast shared memory on the GPU would enable
a two-fold increase of both the WP-basis and method accuracy while
incurring a very mild additional computational cost. The GPU global
memory is typically not saturated in \wpcd{} calculations of \NN{}
scattering calculations. The \wpcd{} method can also be applied to
compute three-nucleon scattering
amplitudes~\cite{PhysRevC.79.064602,Pomerantsev:2014ija}. However,  the limited amount of
global memory might become a constraining factor for such
applications that involve a more complicated Hilbert space basis.

In this study we have focused on the inherent parallelism of the
\wpcd{} method and the opportunities that it offers to benefit from
the use of GPU hardware.
Of course, one could also explore parallelisation of the \wpcd{}
method on a CPU and make parallel use of the fast cores available in
modern CPUs. The efficiency of the GPU \wpcd{} approach is almost
fully determined by the time needed to diagonalize the channel
Hamiltonian and could therefore benefit from the faster CPU clock
speed. Naturally, one could also consider a GPU implementation of the
\mi{} method wherein the \ls{} equation is solved in a parallel
fashion for multiple scattering energies simultaneously. Still, this
would require multiple matrix inversions (one for each interpolation
energy). A simple phase-shift interpolation applied to the \mi{}
method facilitates sufficiently accurate results for \NN{} scattering
observables, at a significantly lower computational cost compared to
explicitly solving the \ls{} equation at every scattering energy of
interest.

\section*{Acknowledgements}
This work was supported by the European Research Council (ERC) under
the European Unions Horizon 2020 research and innovation programme
(Grant agreement No. 758027), and the Swedish Research Council (Grant
No. 2017-04234). The computations were enabled by resources provided
by the Swedish National Infrastructure for Computing (SNIC) at
Chalmers Centre for Computational Science and Engineering (C3SE) and
the National Supercomputer Centre (NSC) partially funded by the
Swedish Research Council.

\appendix

\section{Scattering cross sections and the spin-scattering matrix}
\label{app | scattering theory}
A scattering observable of some operator $\hat{O}$ can generally be written as a trace \cite{Glockle99109},
\begin{equation}
	\langle\hat{O}\rangle = \frac{\text{Tr}\{M\hat{\rho}_iM^\dagger\hat{O}\}}{\text{Tr}\{M\hat{\rho}_iM^\dagger\}} \:,
\end{equation}
where $\hat{\rho}$ is the spin-density matrix, usually represented in a helicity basis, and $M$ is the spin-scattering matrix. 

The helicity-basis projection of the spin-scattering matrix is written as $M_{m_s,m_{s'}}^s$, where $s$ is the conserved total spin and $m_s$ is the corresponding projected spin. These matrices are usually expressed in partial-wave expansions,
\begin{align}
\begin{split}
M_{m_s,m_{s'}}^s = \frac{2\pi}{ip}\sum_{J,l,l'}
&[1-(-1)^{l+s+T_Z}]\sqrt{\frac{2l+1}{4\pi}}\\
&\times Y_{m_s-m_{s'}}^{l'}(\theta,\phi)\\
&\times\langle l',s,m_s-m_{s'},m_{s'}|l',s',j,m_s\rangle \\
&\times\langle l,s,m_s,0|l,s,j,m_s\rangle\\
&\times\left[S^{1J}_{l'l}(p',p) - \delta_{l,l'}\right] \:,
\end{split}
\label{eq | M-matrix PWE}
\end{align}
where the third and fourth rows are Clebsch-Gordan coefficients, $J$
is the total angular momentum, $l$ and $l'$ are the orbital angular
momenta of the inbound and outbound states respectively, $T_Z$ is the
azimuthal isospin projection, $Y_m^l(\theta,\phi)$ is an azimuthal
spherical harmonic, and $S^{1J}_{l'l}(p',p)=\langle p',l',s,J|S|p,l,s,J\rangle$ is the usual scattering matrix.

The on-shell partial-wave-projected $S$-matrix is
related to the on-shell $T(p,p;E)$-matrix element via
\begin{equation}
S^{sJ}_{l'l}(p,p;E) = 1 - 2\pi i T^{sJ}_{l'l}(p,p;E)\:.
\label{eq | S-matrix and T-matrix connection}
\end{equation}
Due to the conservation of probability, the $S$-matrix is unitary, and
can therefore be parametrised by real parameters like phase shifts. In
the Stapp convention \cite{PhysRev.105.302}, the coupled-channel $S$-matrix $S^{1J}_{l'l}$ is
then given by
\begin{equation}
S = \begin{pmatrix}
\cos(2\epsilon_J)e^{2i\delta_{-,J}} & i\sin(2\epsilon_J)e^{i(\delta_{-,J}+\delta_{+,J})} \\
i\sin(2\epsilon_J)e^{i(\delta_{-,J}+\delta_{+,J})} & \cos(2\epsilon_J)e^{2i\delta_{+,J}}
\end{pmatrix} \:,
\end{equation}
where $\delta_{\pm,J}$ refers to $\delta_{l,J}$ for $l=J\pm1$ and
$\epsilon_J$ is the mixing angle. The uncoupled-channel $S$-matrix $S^{0J}_{ll}$ is
given by (for $l=J$)
\begin{equation}
S = e^{2i\delta_{J,J}} \:.
\end{equation}

The angle
$\theta$ denotes the scattering angle between the c.m. inbound and
outbound relative momenta $\bm{p}$ and $\bm{p}'$ respectively, while
$\phi$ is the rotation angle of $\bm{p}'$ around the inbound momentum
$\bm{p}$, but cylindrical symmetry allows us to set $\phi=0$. Using
Eq.~\eqref{eq | M-matrix PWE} we can calculate scattering observables
such as the differential cross section,
\begin{equation}
	\frac{d\sigma}{d\Omega} = \frac{1}{4}\text{Tr}(MM^\dagger)\:.
\end{equation}
However, this approach does not utilise symmetries to reduce the
number of non-contributing terms of $M_{m_s,m_{s'}}^s$. Instead, the
$M$-matrix can be expressed in terms of non-vanishing spin-momentum
products after some consideration of parity conservation, isospin and time-reversal
symmetries, and the Pauli principle. We use the Saclay convention
\cite{Bystricky1978} for these terms,
\begin{align}
	\begin{split}
	M = \frac{1}{2}\big[ (a+b) &+ (a-b)(\bm{\sigma}_1\cdot\bm{n})(\bm{\sigma}_2\cdot\bm{n}) \\
	&+ (c+d)(\bm{\sigma}_1\cdot\bm{m})(\bm{\sigma}_2\cdot\bm{m}) \\
	&+ (c-d)(\bm{\sigma}_1\cdot\bm{l})(\bm{\sigma}_2\cdot\bm{l}) \\
	&+ e((\bm{\sigma}_1+\bm{\sigma}_2)\cdot\bm{m})\big] \:,\end{split}
	\label{eq | M matrix in terms of saclay amplitudes}
\end{align}
where $a,b,c,d$, and $e$ are the Saclay amplitudes, $\bm{\sigma}_i$
are the Pauli spin matrices acting on nucleon $i=1,2$, and where we
define the following unit vectors
\begin{equation}
	\bm{l} \equiv \frac{\bm{p}+\bm{p}'}{|\bm{p}+\bm{p}'|}\:,\:\:
	\bm{m} \equiv \frac{\bm{p}-\bm{p}'}{|\bm{p}-\bm{p}'|}\:,\:\:
	\bm{n} \equiv \frac{\bm{p}\times\bm{p}'}{|\bm{p}\times\bm{p}'|} \:.
	\label{eq | kinematics vectors}
\end{equation}
The Saclay amplitudes are given by the spin-projected matrices as,
\begin{align}
	\begin{split}
	a &= \frac{1}{2}(M^1_{11} + M^1_{0,0} + M^1_{1,-1})\:, \\
	b &= \frac{1}{2}(M^1_{11} + M^0_{0,0} + M^1_{1,-1})\:, \\
	c &= \frac{1}{2}(M^1_{11} - M^0_{0,0} + M^1_{1,-1})\:, \\
	d &= -\frac{1}{\sqrt{2}\sin(\theta)}(M^1_{1,0} + M^1_{0,1})\:, \\
	e &= \frac{i}{2\sqrt{2}}(M^1_{1,0} - M^1_{0,1})\:.
	\end{split}
	\label{eq | saclay amplitudes}
\end{align}
In this parametrisation, the differential cross-section is given by
\begin{equation}
	\frac{d\sigma}{d\Omega} = \frac{1}{2}\left(|a|^2 + |b|^2 + |c|^2 + |d|^2 + |e|^2\right) \:.
\end{equation}
For the results in this paper we used the following expression for the
total cross section:
\begin{equation}
	\sigma_{\text{tot}} = \frac{2\pi}{p}\text{Im}(a+b) \:.
\end{equation}
See Ref.~\cite{Bystricky1978} for a complete account of scattering
observables in the Saclay parametrisation.
\section{The resolvent in a wave-packet basis}
\label{app | resolvent calculation}
The resolvent $\hat{g}(E)$ for the full Hamiltonian
$\hat{h}\equiv\hat{h}_0+\hat{v}$,
\begin{equation}
  \hat{g}(E)\equiv\frac{1}{E-\hat{h} \pm i\epsilon} \:,
\end{equation}
can be calculated analytically in a pseudostate wave-packet basis
$\{|z_i\rangle\}_{i=1}^n$ of the full Hamiltonian. The resolvent is
represented in the basis by (see Eq.~\eqref{eq | definition free wave-packet})
\begin{align}
  \begin{split}
    \langle z_i|\hat{g}(E)|z_j\rangle &= \langle z_i|\frac{1}{E-\hat{h} \pm i\epsilon}|z_j\rangle \\
    &= \frac{1}{\mu\sqrt{N_iN_j}}\int_{\mathcal{D}_i}\int_{\mathcal{D}_j} \diff k'\diff k \\
    &\hspace{2cm}\times\frac{k'k\sqrt{k'k}\langle \psi_{k'}^{(+)}|\psi_k^{(+)}\rangle}{E- \frac{k'^2}{2\mu} \pm i\epsilon} \:,
  \end{split}
\end{align}
where $\mu=\frac{m_N}{2}$. Note that we set the weight function $f(k)=\sqrt{\frac{k}{\mu}}$ and
normalisation $N_i=\Delta\mathcal{E}_i$ as these are energy wave-packets (from
the diagonalisation of $\hat{h}$). Using $\langle
\psi_{k'}^{(+)}|\psi_k^{(+)}\rangle=\frac{\delta(k'-k)}{k'k}$, this becomes
\begin{equation}
  \langle z_i|\hat{g}(E)|z_j\rangle = \frac{\delta_{ij}}{\mu N_i}\int_{\mathcal{D}_i} \diff k \:  \frac{k}{E-\frac{k^2}{2\mu} \pm i\epsilon} \:,
\end{equation}
where we have introduced the Kronecker delta $\delta_{ij}$ since
$\langle \psi_{k'}^{(+)}|\psi_k^{(+)}\rangle=0\:\forall\:\mathcal{D}_i\neq\mathcal{D}_j$. For
positive energies, where $E = \frac{p^2}{2\mu}$ and where $p$ is the
on-shell c.m. momentum, we get
\begin{equation}
  \langle z_i|\hat{g}(E)|z_j\rangle = \frac{2\delta_{ij}}{D_i}\int_{\mathcal{D}_i} \diff k \:  \frac{k}{p^2-k^2 \pm i\epsilon} \:,
\end{equation}
If $E\notin\mathcal{D}_i$, we take the limit
$\epsilon\rightarrow0$ and solve the integral to find
\begin{align}
  \begin{split}
    \langle z_i|\hat{g}(E)|z_j\rangle &= \frac{2\delta_{ij}}{N_i}\int_{\mathcal{D}_i} \diff k \:  \frac{k}{p^2-k^2} \\
    &= \frac{\delta_{ij}}{N_i}\left[-\ln\left|\frac{k^2}{p^2}-1\right|\right]_{k_{i}}^{k_{i+1}} \\
    &= \frac{\delta_{ij}}{N_i}\ln\left| \frac{E+\mathcal{E}_{i+1}}{E+\mathcal{E}_{i}} \right| \:,
  \end{split}
\end{align}
where $p\notin\mathcal{D}_i$, and $\mathcal{E}_i$ and $\mathcal{E}_{i+1}$ is the lower and
upper boundary of $\mathcal{D}_i$ expressed in energy,
respectively. If $E\in\mathcal{D}_i$, then we have a simple pole at
$p=k$. The pole-integration is done using the infinitesimal complex
rotation $\pm i\epsilon$ together with the residue theorem, giving
\begin{align}
  \begin{split}
    \langle z_i|\hat{g}(E)|z_j\rangle = \frac{\delta_{ij}}{N_i}\bigg[&\ln\left| \frac{E+\mathcal{E}_{i+1}}{E+\mathcal{E}_{i}} \right| \\ &- i\pi(\theta(E-\mathcal{E}_i) - \theta(E-\mathcal{E}_{i+1}))\bigg] \:,
  \end{split}
\end{align}
where $\theta$ is the Heaviside step-function. The derivation of
the resolvent expressed in a momentum wave-packet representation
follows a similar procedure. In that case, it is possible to use
momentum wave-packets where $f(p)=1$ and $N_i=\sqrt{k_{i+1} - k_i}$, in which
case the derivation above changes a little, see \cite{RUBTSOVA2015613}.\\

Energy averaging of the resolvent is done by integrating the resolvent
with respect to $E$, in the bin $E\in\mathcal{D}_k$, divided by the
bin width $\Delta\mathcal{E}_k$. We introduce the denotation $\bar{g}_{ij}^k$ to
reflect this. The derivation is straightforward:
\begin{align}
  \begin{split}
    \bar{g}_{ij}^k &\equiv \frac{1}{\Delta\mathcal{E}_k}\int_{\mathcal{D}_k} \diff E\:\langle z_i|\hat{g}(E)|z_j\rangle \\
    &= \frac{\delta_{ij}}{\Delta\mathcal{E}_k N_i}\int_{\mathcal{D}_k}\diff E\: \left[\ln\left| \frac{E+\mathcal{E}_{i+1}}{E+\mathcal{E}_{i}} \right| - i\pi\delta_{ik}\right] \\
    &= \frac{\delta_{ij}}{\Delta\mathcal{E}_k \Delta\mathcal{E}_{i}}W_{ki} - \frac{i\pi}{\Delta\mathcal{E}_k}\delta_{ik} \:,
  \end{split}
  \label{eq | energy averaged resolvent}
\end{align}
where we used $N_i=\Delta\mathcal{E}_{i}$, and
\begin{equation}
  W_{ki} \equiv \sum_{k'=k}^{k+1}\sum_{i'=i}^{i+1}(-1)^{k-k'+i-i'}\left[\mathcal{E}_{k'}- \mathcal{E}_{i'}\right]\ln\left|\mathcal{E}_{k'}- \mathcal{E}_{i'}\right|\:.
\end{equation}
as presented in \cite{RUBTSOVA2015613}.
\begin{figure}
  \centering
  \includegraphics[width=0.7\columnwidth]{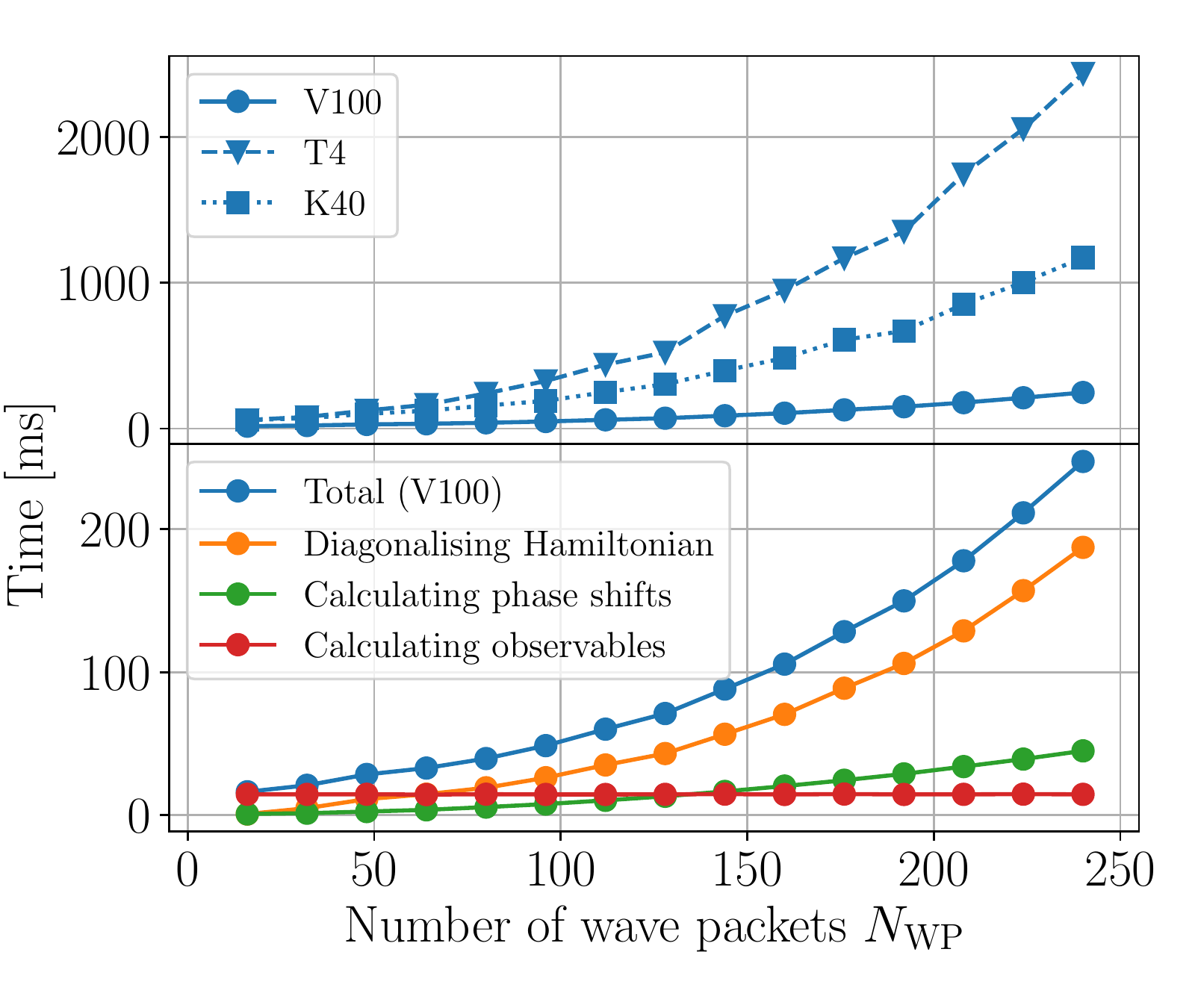}
  \caption{Top: total time used in calculating on-shell $T$-matrix elements and
    extracting phase shifts and observables, shown for three different Nvidia
    GPUs of the Tesla line; T4 (orange), V100 (blue), and K40
    (green). Bottom: time decomposition of V100 total time, with focus
    on key parts of calculating \NN{} scattering using the \wpcd{}
    method. The green line includes the time spent solving the \ls{} equation.
    \label{fig | gpu real profilies}%
    }
\end{figure}
\section{GPU code}
\label{app | gpu code}
The code for GPU-utilisation was written to make use of the CUDA
interface, which is developed and maintained by the Nvidia
Corporation. CUDA allows for efficient utilisation of a Nvidia GPU
using high-level programming languages such as C and C++, Python, or
Fortran. For numerically demanding linear algebra operations we made
use of the CUDA-libraries cuBLAS\cite{cublas} and
cuSOLVER\cite{cusolver}. These libraries are very similar in use to
BLAS and LAPACK---two standard libraries for linear algebra on the
CPU.
	
Diagonalising the full Hamiltonian and solving the \ls{} equation are the
most numerically demanding parts of the steps presented in
Sec.~\ref{sec:WPCD}. These steps are therefore solved on the GPU. The
ability to work on several channels simultaneously---this being the
advantage of the GPU---is made possible by using CUDA's
\emph{batched}-routines. These are pre-written routines that maximise
the efficiency of the GPU to solve several linear algebra problems
simultaneously for small matrix-sizes (typically less than
$1000\times 1000$ in matrix dimension). Such efficient parallelism is
generally difficult to achieve ``by hand''due to the massive load on
GPU-memory read-write accesses.
	
The importance of efficient memory use is made apparent in
Fig.~\ref{fig | gpu real profilies} where we show the computation time spent by
three different Nvidia GPUs (top panel) and the decomposition of time used for the
Nvidia V100 GPU (bottom panel). We see that the majority of the total time is spent
on the Hamiltonian diagonalisation. A large fraction of the
diagonalisation task is spent on memory accesses as part of the Jacobi
method.  The three GPUs: V100, T4, and K40,  have differences in memory
technology~\cite{jia2019dissecting} which is the main reason for the
observed  differences in the performance.
	
We used\texttt{cublas<t>gemmStridedBatched} to calculate the
$VC$-matrix product in Eq.~\eqref{eq | on-shell T-mat element in
  fWP}. This calculates a matrix-product on the form $C \leftarrow
\alpha AB + \beta C$, where $C$ is overwritten by the right-hand
side. This setup is standard for BLAS \texttt{gemm}-routines. Here,
$\alpha$ and $\beta$ are scalars, while $A$, $B$, and $C$ are sets of
matrices stored congruently in three arrays, i.e. they are
``batched''.
	
To diagonalise the Hamiltonians we used
\texttt{cusolverDn<t>syevjBatched}. This routine utilises the parallel
cyclic-order Jacobi method, as was briefly discussed in Sec.~\ref{sec
  | WPCD complexity}, to diagonalise batches of matrices
simultaneously.
	
Lastly, solving Eq.~\eqref{eq | LS WP sum} was done using a
custom-written function, referred to as a ``kernel'' in CUDA. The
advantage of using the GPU for this task comes with the energy-dependence in
the resolvent. We can calculate all on-shell $T$-matrix elements
simultaneously following a single batched Hamiltonian diagonalisation.

\section*{References}
\bibliographystyle{unsrt}
\bibliography{bibliography}

\begin{thebibliography}{10}

\bibitem{doi:10.1146/annurev.nucl.52.050102.090637}
P.~F. Bedaque and U.~van Kolck.
\newblock Effective field theory for few-nucleon systems.
\newblock {\em Annu. Rev. Nucl. Part. Sci.}, 52(1):339--396, 2002.

\bibitem{RevModPhys.81.1773}
E.~Epelbaum, H.-W. Hammer, and Ulf-G. Mei\ss{}ner.
\newblock Modern theory of nuclear forces.
\newblock {\em Rev. Mod. Phys.}, 81:1773--1825, 12 2009.

\bibitem{MACHLEIDT20111}
R.~Machleidt and D.~R. Entem.
\newblock Chiral effective field theory and nuclear forces.
\newblock {\em Phys. Rep.}, 503(1):1 -- 75, 2011.

\bibitem{RevModPhys.92.025004}
H.-W. Hammer, Sebastian K\"onig, and U.~van Kolck.
\newblock Nuclear effective field theory: Status and perspectives.
\newblock {\em Rev. Mod. Phys.}, 92:025004, Jun 2020.

\bibitem{Wesolowski_2016}
S.~Wesolowski, N.~Klco, R.~J. Furnstahl, D.~R. Phillips, and A.~Thapaliya.
\newblock Bayesian parameter estimation for effective field theories.
\newblock {\em J. Phys. G}, 43(7):074001, may 2016.

\bibitem{PhysRevC.68.041001}
D.~R. Entem and R.~Machleidt.
\newblock Accurate charge-dependent nucleon-nucleon potential at fourth order
  of chiral perturbation theory.
\newblock {\em Phys. Rev. C}, 68:041001, Oct 2003.

\bibitem{PhysRevX.6.011019}
B.~D. Carlsson, A.~Ekstr\"om, C.~Forss\'en, D.~Fahlin Str\"omberg, G.~R.
  Jansen, O.~Lilja, M.~Lindby, B.~A. Mattsson, and K.~A. Wendt.
\newblock Uncertainty analysis and order-by-order optimization of chiral
  nuclear interactions.
\newblock {\em Phys. Rev. X}, 6:011019, 2 2016.

\bibitem{Reinert:2018tb}
P.~Reinert, H.~Krebs, and E.~Epelbaum.
\newblock {Semilocal momentum-space regularized chiral two-nucleon potentials
  up to fifth order}.
\newblock {\em Eur. Phys. J. A}, 54(5):86, 2018.

\bibitem{PhysRevC.91.024003}
M.~Piarulli, L.~Girlanda, R.~Schiavilla, R.~Navarro P\'erez, J.~E. Amaro, and
  E.~Ruiz Arriola.
\newblock Minimally nonlocal nucleon-nucleon potentials with chiral two-pion
  exchange including $\ensuremath{\Delta}$ resonances.
\newblock {\em Phys. Rev. C}, 91:024003, Feb 2015.

\bibitem{PhysRevC.88.064002}
R.~Navarro P\'erez, J.~E. Amaro, and E.~Ruiz Arriola.
\newblock Coarse-grained potential analysis of neutron-proton and proton-proton
  scattering below the pion production threshold.
\newblock {\em Phys. Rev. C}, 88:064002, Dec 2013.

\bibitem{Frame:2017fah}
Dillon Frame, Rongzheng He, Ilse Ipsen, Daniel Lee, Dean Lee, and Ermal Rrapaj.
\newblock {Eigenvector continuation with subspace learning}.
\newblock {\em Phys. Rev. Lett.}, 121(3):032501, 2018.

\bibitem{Konig:2019adq}
S.~K\"onig, A.~Ekstr\"om, K.~Hebeler, D.~Lee, and A.~Schwenk.
\newblock {Eigenvector Continuation as an Efficient and Accurate Emulator for
  Uncertainty Quantification}.
\newblock {\em Phys. Lett. B}, 810:135814, 2020.

\bibitem{Furnstahl:2020abp}
R.~J. Furnstahl, A.~J. Garcia, P.~J. Millican, and X.~Zhang.
\newblock {Efficient emulators for scattering using eigenvector continuation}.
\newblock {\em Phys. Lett. B}, 809:135719, 2020.

\bibitem{Melendez:2021lyq}
J.~A. Melendez, C.~Drischler, A.~J. Garcia, R.~J. Furnstahl, and Xilin Zhang.
\newblock {Fast \& accurate emulation of two-body scattering observables
  without wave functions}.
\newblock 6 2021.

\bibitem{zhang2021fast}
Xilin Zhang and R.~J. Furnstahl.
\newblock Fast emulation of quantum three-body scattering, 2021.

\bibitem{10.5555/1162254}
Carl~Edward Rasmussen and Christopher K.~I. Williams.
\newblock {\em Gaussian Processes for Machine Learning (Adaptive Computation
  and Machine Learning)}.
\newblock The MIT Press, 2005.

\bibitem{Haftel1970}
M.~I. Haftel and F.~Tabakin.
\newblock Nuclear saturation and the smoothness of nucleon-nucleon potentials.
\newblock {\em Nucl. Phys. A}, 158(1):1 -- 42, 1970.

\bibitem{RUBTSOVA2015613}
O.~A. Rubtsova, V.~I. Kukulin, and V.~N. Pomerantsev.
\newblock Wave-packet continuum discretization for quantum scattering.
\newblock {\em Ann. Phys. (N. Y.)}, 360:613--654, 2015.

\bibitem{Brynjarsdottir:2014cb}
J.~Brynjarsd{\'o}ttir and A.~O'Hagan.
\newblock {Learning about physical parameters: the importance of model
  discrepancy}.
\newblock {\em Inverse Probl.}, 30(11):114007--25, October 2014.

\bibitem{PhysRevC.96.024003}
J.~A. Melendez, S.~Wesolowski, and R.~J. Furnstahl.
\newblock Bayesian truncation errors in chiral effective field theory:
  Nucleon-nucleon observables.
\newblock {\em Phys. Rev. C}, 96:024003, 8 2017.

\bibitem{lh1}
L.~Hulth\'en.
\newblock {\em Kungl. Fysiogr. S\"allsk. i Lund F\"orhandl.}, 14, 1944.

\bibitem{lh2}
L.~Hulth\'en.
\newblock {\em Arkiv Mat. Astron. Fysik}, 35A, 1948.

\bibitem{PhysRev.74.1763}
W.~Kohn.
\newblock Variational methods in nuclear collision problems.
\newblock {\em Phys. Rev.}, 74:1763--1772, Dec 1948.

\bibitem{PhysRev.141.1468}
C.~Schwartz.
\newblock Application of the schwinger variational principle for scattering.
\newblock {\em Phys. Rev.}, 141:1468--1470, Jan 1966.

\bibitem{doi:10.1002/prop.19720200502}
J.~L. Basdevant.
\newblock The padé approximation and its physical applications.
\newblock {\em Fortschr. Phys.}, 20(5):283--331, 1972.

\bibitem{Erkelenz:1971caz}
K.~Erkelenz, R.~Alzetta, and K.~Holinde.
\newblock {Momentum space calculations and helicity formalism in nuclear
  physics}.
\newblock {\em Nucl. Phys. A}, 176:413--432, 1971.

\bibitem{10.1093/imanum/12.1.1}
J.~J.~du Croz and N.~J. Higham.
\newblock {Stability of Methods for Matrix Inversion}.
\newblock {\em IMA J. Numer. Anal.}, 12(1):1--19, 1 1992.

\bibitem{Glockle99109}
W.~Gl{\"o}ckle.
\newblock {\em The quantum mechanical few-body problem}.
\newblock Springer Science \& Business Media, 2012.

\bibitem{CARBONELL201455}
J.~Carbonell, A.~Deltuva, A.~C. Fonseca, and R.~Lazauskas.
\newblock Bound state techniques to solve the multiparticle scattering problem.
\newblock {\em Prog. Part. Nucl. Phys.}, 74:55 -- 80, 2014.

\bibitem{Heller19751222}
E.~J. Heller.
\newblock Theory of $j$-matrix green's functions with applications to atomic
  polarizability and phase-shift error bounds.
\newblock {\em Phys. Rev. A}, 12:1222--1231, Oct 1975.

\bibitem{Winick1978910}
J.~R. Winick and W.~P. Reinhardt.
\newblock Moment $t$-matrix approach to ${e}^{+}$-h scattering. i. angular
  distribution and total cross section for energies below the pickup threshold.
\newblock {\em Phys. Rev. A}, 18:910--924, Sep 1978.

\bibitem{Winick1978925}
J.~R. Winick and W.~P. Reinhardt.
\newblock Moment $t$-matrix approach to ${e}^{+}$-h scattering. ii. elastic
  scattering and total cross section at intermediate energies.
\newblock {\em Phys. Rev. A}, 18:925--934, Sep 1978.

\bibitem{Heller19732946}
E.~J. Heller, T.~N. Rescigno, and W.~P. Reinhardt.
\newblock Extraction of scattering information from fredholm determinants
  calculated in an ${L}^{2}$ basis: A chebyschev discretization of the
  continuum.
\newblock {\em Phys. Rev. A}, 8:2946--2951, Dec 1973.

\bibitem{Corcoran1977}
C.~T. Corcoran and P.~W. Langhoff.
\newblock Moment‐theory approximations for nonnegative spectral densities.
\newblock {\em J. Math. Phys.}, 18(4):651--657, 1977.

\bibitem{Rubtsova20072025}
O.~A. Rubtsova and V.~I. Kukulin.
\newblock Wave-packet discretization of a continuum: Path toward practically
  solving few-body scattering problems.
\newblock {\em Phys. At. Nucl.}, 70(12):2025--2045, Dec 2007.

\bibitem{kukulin2007wave}
V.~I. Kukulin, V.~N. Pomerantsev, and O.~A. Rubtsova.
\newblock Wave-packet continuum discretization method for solving the
  three-body scattering problem.
\newblock {\em Theor. Math. Phys.}, 150(3):403--424, Mar 2007.

\bibitem{PhysRevC.79.064602}
O.~A. Rubtsova, V.~N. Pomerantsev, and V.~I. Kukulin.
\newblock Quantum scattering theory on the momentum lattice.
\newblock {\em Phys. Rev. C}, 79:064602, 6 2009.

\bibitem{Kukulin2003404}
V.~I. Kukulin and O.~A. Rubtsova.
\newblock Discrete quantum scattering theory.
\newblock {\em Theor. Math. Phys.}, 134(3):404--426, Mar 2003.

\bibitem{PhysRevC.94.024328}
H.~M\"uther, O.~A. Rubtsova, V.~I. Kukulin, and V.~N. Pomerantsev.
\newblock Discrete wave-packet representation in nuclear matter calculations.
\newblock {\em Phys. Rev. C}, 94:024328, 8 2016.

\bibitem{TheorMathPhys145.1711.1726}
V.~I. Kukulin and O.~A. Rubtsova.
\newblock {Solving the Charged-Particle Scattering Problem by Wave Packet
  Continuum Discretization}.
\newblock {\em Theor. Math. Phys.}, 145:1711–--1726, Dec 2005.

\bibitem{PhysRevC.81.064003}
O.~A. Rubtsova, V.~I. Kukulin, V.~N. Pomerantsev, and A.~Faessler.
\newblock New approach toward a direct evaluation of the multichannel
  multienergy $s$ matrix without solving the scattering equations.
\newblock {\em Phys. Rev. C}, 81:064003, Jun 2010.

\bibitem{cublas}
Dense linear algebra on gpus.
\newblock \url{https://developer.nvidia.com/cublas}, 2021.
\newblock Accessed: 2021-03-22.

\bibitem{cusolver}
Cuda toolkit documentation.
\newblock \url{https://docs.nvidia.com/cuda/cusolver/index.html}, 2021.
\newblock Accessed: 2021-03-22.

\bibitem{GoluVanl96}
G.~H. Golub and C.~F. Van~Loan.
\newblock {\em Matrix Computations}.
\newblock The Johns Hopkins University Press, Baltimore, MD, USA, third
  edition, 1996.

\bibitem{Pourzandi1994}
M.~Pourzandi and B.~Tourancheau.
\newblock A parallel performance study of jacobi-like eigenvalue solution.
\newblock {\em Technical Report}, 05 1994.

\bibitem{ABDELFATTAH2020188}
A.~Abdelfattah, S.~Tomov, and J.~Dongarra.
\newblock Matrix multiplication on batches of small matrices in half and
  half-complex precisions.
\newblock {\em J. Parallel Distrib. Comput.}, 145:188 -- 201, 2020.

\bibitem{BAE20142230}
S.~E. Bae, T.-W. Shinn, and T.~Takaoka.
\newblock A faster parallel algorithm for matrix multiplication on a mesh
  array.
\newblock {\em Procedia Comput. Sci.}, 29:2230 -- 2240, 2014.
\newblock 2014 International Conference on Computational Science.

\bibitem{PhysRevLett.110.192502}
A.~Ekstr\"om, G.~Baardsen, C.~Forss\'en, G.~Hagen, M.~Hjorth-Jensen, G.~R.
  Jansen, R.~Machleidt, W.~Nazarewicz, T.~Papenbrock, J.~Sarich, and S.~M.
  Wild.
\newblock Optimized chiral nucleon-nucleon interaction at
  next-to-next-to-leading order.
\newblock {\em Phys. Rev. Lett.}, 110:192502, 5 2013.

\bibitem{osti_4434712}
N.~Levinson.
\newblock On the uniqueness of the potential in a schrodinger equation for a
  given asymptotic phase.
\newblock {\em Kgl. Danske Videnskab Selskab. Mat. Fys. Medd.}, 25(9), 1 1949.

\bibitem{PhysRevLett.49.255}
P.~W. Lisowski, R.~E. Shamu, G.~F. Auchampaugh, N.~S.~P. King, M.~S. Moore,
  G.~L. Morgan, and T.~S. Singleton.
\newblock Search for resonance structure in the $\mathrm{np}$ total cross
  section below 800 mev.
\newblock {\em Phys. Rev. Lett.}, 49:255--259, Jul 1982.

\bibitem{PhysRevC.63.044608}
W.~P. Abfalterer, F.~B. Bateman, F.~S. Dietrich, R.~W. Finlay, R.~C. Haight,
  and G.~L. Morgan.
\newblock Measurement of neutron total cross sections up to 560 mev.
\newblock {\em Phys. Rev. C}, 63:044608, Mar 2001.

\bibitem{Pomerantsev:2014ija}
V.~N. Pomerantsev, V.~I. Kukulin, and O.~A. Rubtsova.
\newblock {New general approach in few-body scattering calculations: Solving
  discretized Faddeev equations on a graphics processing unit}.
\newblock {\em Phys. Rev. C}, 89(6):064008, 2014.

\bibitem{PhysRev.105.302}
H.~P. Stapp, T.~J. Ypsilantis, and N.~Metropolis.
\newblock Phase-shift analysis of 310-mev proton-proton scattering experiments.
\newblock {\em Phys. Rev.}, 105:302--310, 1 1957.

\bibitem{Bystricky1978}
J.~Bystricky, F.~Lehar, and P.~Winternitz.
\newblock Formalism of nucleon-nucleon elastic scattering experiments.
\newblock {\em J. Phys. France}, 39(1):1--32, 1978.

\bibitem{jia2019dissecting}
Z.~Jia, M.~Maggioni, J.~Smith, and D.~P. Scarpazza.
\newblock Dissecting the nvidia turing t4 gpu via microbenchmarking, 2019.

\end{thebibliography}

\end{document}